\documentclass[twocolumn]{aastex62}

\usepackage[caption=false]{subfig}
\usepackage{float}
\usepackage{graphicx}	
\usepackage{amsmath}	
\usepackage{amssymb}
\usepackage{xcolor}
\usepackage{booktabs}
\usepackage{enumerate}
\usepackage{dirtytalk}
\usepackage{float}
\usepackage{enumitem}
\usepackage{booktabs}
\usepackage{lipsum} 
\usepackage{multirow}
\usepackage{makecell}
\usepackage[english]{babel}

\usepackage{multirow}
\usepackage[para]{footmisc}
\usepackage{footnote}
\usepackage[normal]{threeparttable}

\graphicspath{{./}}

\received{26 June 2020}
\accepted{28 April 2021}
\published{26 August 2021}

\submitjournal{The Astrophysical Journal}

\shorttitle{Hycean Habitability and Biosignatures}
\shortauthors{Madhusudhan et al.}

\begin{document}

\title{Habitability and Biosignatures of Hycean Worlds}

\correspondingauthor{Nikku Madhusudhan (nmadhu@ast.cam.ac.uk)}


\author[0000-0002-4869-000X]{Nikku Madhusudhan}
\affil{Institute of Astronomy, University of Cambridge, Madingley Road, Cambridge CB3 0HA, UK}

\author[0000-0002-4487-5533]{Anjali A. A. Piette}
\affil{Institute of Astronomy, University of Cambridge, Madingley Road, Cambridge CB3 0HA, UK}

\author[0000-0001-6839-4569]{Savvas Constantinou}
\affil{Institute of Astronomy, University of Cambridge, Madingley Road, Cambridge CB3 0HA, UK}

\begin{abstract}
We investigate a new class of habitable planets composed of water-rich interiors with massive oceans underlying H$_2$-rich atmospheres, referred to here as Hycean worlds. With densities between those of rocky super-Earths and more extended mini-Neptunes, Hycean planets can be optimal candidates in the search for exoplanetary habitability and may be abundant in the exoplanet population. We investigate the bulk properties (masses, radii, and temperatures),  potential for habitability, and observable biosignatures of Hycean planets. We show that Hycean planets can be significantly larger compared to previous considerations for habitable planets, with radii as large as 2.6 $R_\oplus$ (2.3 $R_\oplus$) for a mass of 10 $M_\oplus$ (5 $M_\oplus$). We construct the Hycean habitable zone (HZ), considering stellar hosts from late M to sun-like stars, and find it to be significantly wider than the terrestrial-like HZ. While the inner boundary of the Hycean HZ corresponds to equilibrium temperatures as high as $\sim$500~K for late M dwarfs, the outer boundary is unrestricted to arbitrarily large orbital separations. Our investigations include tidally locked `Dark Hycean' worlds that permit habitable conditions only on their permanent nightsides and `Cold Hycean' worlds that see negligible irradiation. Finally, we investigate the observability of possible biosignatures in Hycean atmospheres. We find that a number of trace terrestrial biomarkers which may be expected to be present in Hycean atmospheres would be readily detectable using modest observing time with the James Webb Space Telescope (JWST). We identify a sizable sample of nearby  potential Hycean planets that can be ideal targets for such observations in search of exoplanetary biosignatures. 
\end{abstract}

\keywords{Exoplanets --- Habitable planets --- Exoplanet atmospheres --- Radiative transfer --- Planetary interior --- Biosignatures --- Transmission spectroscopy}

\section{Introduction} 
\label{sec:intro}

Of the thousands of exoplanets known today , the vast majority are low-mass planets with sizes of 1-4 $R_\oplus$, between the terrestrial planets and ice giants of the solar system \citep{howard2012,fressin2013,petigura2013,fulton2018,hardegree-ullman2020}. With no analogs in the solar system, these planets are variedly classed as super-Earths or mini-Neptunes depending on fiducial inferences about their bulk compositions based on their densities \citep[e.g.,][]{valencia2007,rogers2011,lopez2012,lopez2014,rogers2015}. Recent surveys are discovering a number of low-mass planets in the habitable zones (HZs) of their host stars, notably exoplanets in HZs of nearby M stars \citep{tarter2007,mulders2015}, e.g., TRAPPIST-1 \citep{gillon2017}, Proxima Cen \citep{anglada-escude2016}, K2-18 \citep{foreman-mackey2015,montet2015} and LHS~1140 \citep{dittmann2017}. Their nearby and bright host stars make such planets conducive for detailed characterization. In particular, establishing the habitability of such planets requires characterization of their atmospheres, paving the way for potential biosignature detections \citep[e.g.,][]{seager2013a,seager2016,kaltenegger2017, meadows2018}. 
Tremendous progress has been made in the characterization of exoplanetary atmospheres \citep[e.g.,][]{seager2010,birkby2018,kreidberg2018,madhu2018,madhu2019}. The smallest planets whose atmospheres have been characterized to date are mini-Neptunes where H$_2$O features have been observed in transmission spectra in the near-infrared with the Hubble Space Telescope \citep[HST; e.g.][]{benneke2019a,benneke2019,tsiaras2019}. Atmospheric observations of HZ terrestrial exoplanets are still very challenging. The detection of an atmospheric signature for an Earth-like habitable planet orbiting a sun-like star remains a difficult goal \citep{Kaltenegger2009,arnold2014,feng2018}. However, HZ rocky exoplanets orbiting M Dwarfs are more accessible. Theoretical studies show that upcoming large facilities such as the James Webb Space Telescope (JWST) and the Extremely Large Telescope (ELT) will have the capability to detect potential atmospheric biosignatures in such planets, but with significant investment of observing time \citep{snellen2013,rodler2014, barstow2016, lustig-yaeger2019}. These challenges call for new, more accessible, avenues to pursue the search for habitable exoplanets and biosignatures. The possibility of exoplanetary habitability depends on 
both the atmospheric and internal structure of the planet, which governs the surface conditions, presence of oceans, and potential for life. 

The interiors of planets in the low-mass regime can span a diverse range of compositions. These range from predominantly rocky super-Earths \citep[e.g.,][]{fortney2007,seager2007,valencia2007,elkins-tanton2008,wagner2011,zeng2013} to mini-Neptunes akin to ice giants in the solar system, i.e., with a significant mass fraction in volatile ices and the H$_2$/He envelope \citep[e.g.,][]{rogers2010a,nettelmann2011,rogers2011,valencia2013}. Previous studies have also investigated the possibility of water worlds, with substantial mass fractions of H$_2$O  \citep{kuchner2003,leger2004,selsis2007a,sotin2007,marcus2010,nettelmann2011,alibert2014, zeng2014,thomas2016}. 

\citet{leger2004} proposed the possibility of habitable ocean worlds with atmospheres of terrestrial-like composition, e.g., dominated by N$_2$, H$_2$O, and CO$_2$. Various other studies have investigated the habitability of such ocean worlds \citep[e.g.,][]{kitzmann2015,noack2017,kite2018,ramirez2018}. Recent studies have also  investigated such water-rich planets over a wide range of temperatures \citep[e.g.,][]{zeng2014,thomas2016,zeng2019,mousis2020}, and show that such a composition may explain the masses and radii of a sizable fraction of mini-Neptunes \citep[e.g.,][]{zeng2019,mousis2020}. In particular, a subset of temperate mini-Neptunes could allow for a liquid water surface underneath an H$_2$/He atmosphere, making them conducive for habitability as recently suggested for the HZ planet K2-18~b \citep{madhu2020,piette2020}. 

Traditionally, the HZ around a star is defined by the requirement of liquid water on the surface of an Earth-like rocky planet \citep[e.g.,][]{hart1978,kasting1993,forget1998,kasting2003, selsis2007,selsis2008,forget2013, yang2013, zsom2013,kaltenegger2017,kopparapu2018,meadows2018book}. Typically, the atmospheric composition is considered  to be dominated by a combination of N$_2$, O$_2$, CO$_2$, and H$_2$O, similar to atmospheres of solar system terrestrial planets \citep[e.g.,][]{kasting1993,kopparapu2016}. In this case, the inner edge of the HZ is restricted by the runaway greenhouse effect and/or escape of water from the atmosphere \citep[e.g.,][]{rasool1970,hart1978,Abe1988,kasting1988,leconte2013b,wolf2015,ribas2016,bolmont2017,kopparapu2017}. Conversely, the outer edge of the HZ is generally limited by CO$_2$ condensation preventing the greenhouse warming needed to sustain liquid H$_2$O  \citep[e.g.,][]{kasting1993,turbet2016,turbet2018}. 
  
Some studies have also investigated the habitability of rocky exoplanets with H$_2$-rich atmospheres \citep{stevenson1999,pierrehumbert2011,wordsworth2012,koll2019}. For example, \citet{stevenson1999} suggests habitable conditions on Earth-like or smaller rocky planets or planetary embryos in interstellar space with no stellar insolation. \citet{pierrehumbert2011} consider rocky planets with H$_2$/He atmospheres and low stellar insolation beyond the traditional HZ, showing that habitable conditions on such planets may be possible out to 1.5 AU for M dwarf stars and 10 AU for G dwarfs. Conversely, \citet{koll2019} explore the inner edge of the HZ for Earth-like planets with H$_2$-rich atmospheres orbiting Sun-like stars, in particular the impact of greenhouse warming due to H$_2$O. While it has been suggested that significantly larger mini-Neptunes with H$_2$/He atmospheres could also be potentially habitable \citep{madhu2020}, their implications for the HZ have not been fully explored. 

Ultimately, establishing the presence of life on a habitable exoplanet requires the detection of reliable biomarkers in its atmosphere. The prominent biomarkers that have traditionally been considered based on the Earth's atmosphere are O$_2$, O$_3$, CH$_4$, and N$_2$O \citep[e.g.,][]{owen1980,leger1993,sagan1993,desmarais2002,catling2018,schwieterman2018}. While these molecules are predominantly a result of life on Earth, they have also been proposed to be contributed, albeit in small amounts, by abiogenic sources \citep{etiope2013,meadows2017,catling2018,schwieterman2018}.
At the same time, a number of less abundant molecules are also known to have originated from metabolic processes in Earth's biosphere \citep[see, e.g., reviews by][]{pilcher2003,catling2018,schwieterman2018}. These include a number of organosulfur compounds such as dimethysulfide (DMS), dimethyldisulfide (DMDS), methanethiol (CH$_3$SH), and carbonylsulfide (OCS), whose origins in Earth's biosphere have been extensively studied \citep[e.g.,][]{andrea1983, cline1983, vairavamurthy1985, cooper1987, bates1992, pilcher2003, visscher2003}. 

The feasibility of such trace biomarkers in exoplanetary atmospheres has been explored in various recent studies \citep[e.g.,][]{pilcher2003,segura2005,domagal-goldman2011, seager2013a,seager2013b} alongside the more traditional and dominant molecules such as O$_2$, O$_3$ and CH$_4$. It is well known that life originated on the early Earth  before O$_2$ and O$_3$ became abundant in the atmosphere \citep[e.g.,][]{schopf1983,holland1984,arnold2004,bekker2004,stolper2010,lyons2014}, implying that a nondetection of O$_2$/O$_3$ does not rule out the possibility of life on an exoplanet \citep{domagal-goldman2011}. In particular, it has been shown that molecules such as DMS, DMDS, OCS, CH$_3$Cl, and N$_2$O can be prevalent in the atmospheres of terrestrial exoplanets with similar strengths of biogenic sources to those on Earth under different stellar hosts and atmospheric conditions \citep{segura2005,domagal-goldman2011,seager2013b,seager2016}. 

It is also known that the same terrestrial biomarkers can survive in H$_2$-rich atmospheres. Microorganisms on Earth are known to survive in H$_2$-rich environments \citep{stevens1995,freund2002,gregory2019} including conditions with up to $\sim$88\% H$_2$ concentrations in natural environments \citep{gregory2019}, and even 100\% in laboratory conditions \citep{seager2020}. In the reducing conditions of the early Earth, molecules such as DMS, DMDS, OCS, and CS$_2$ may have been prominent biosignatures \citep[e.g.,][]{pilcher2003,domagal-goldman2011}. Several of the typical biosignatures in Earth's present atmosphere are either not very abundant (e.g., O$_2$ and O$_3$) or not uniquely identifiable as biosignatures (e.g., CH$_4$) in H$_2$-rich atmospheres \citep{seager2013b,seager2016}. In the latter case, CH$_4$ can be a natural carrier of carbon in H$_2$-rich atmospheres, and its abundance could dwarf that produced by biological sources. Reliable and observable biosignatures in H$_2$-rich environments are, therefore, expected to be those gases released from secondary metabolic processes of microorganisms as discussed above, e.g., CH$_3$Cl, DMS, CS$_2$, N$_2$O, and OCS \citep{seager2013b,seager2016}. All these molecules are expected to be present in trace quantities at the $\sim$1 part per million by volume (ppmv) level, but they are expected to be detectable in transmission spectroscopy with JWST for rocky super-Earths with H$_2$-rich atmospheres \citep{seager2013b,seager2016}. 

In this work, we focus on planets with a large fraction of their mass in H$_2$O and with H$_2$-rich atmospheres. Such planets have generally been classified as ``mini-Neptunes'', which are typically assumed to have radii below that of Neptune, i.e., 4 $R\oplus$, but larger than $\sim$1.6-2 $R_\oplus$ \citep{borucki2011,lopez2014,rogers2015}. These objects are smaller than ice giants but too large to have predominantly rocky interiors like super-Earths \citep{rogers2011,lopez2014}. Past explorations of mini-Neptune interiors have found that in some cases the pressure and temperature beneath the H$_2$-rich envelope would be too high to allow for habitability, e.g., in the case of GJ~1214~b \citep[e.g.,][]{rogers2010,nettelmann2011}. However, it has recently been shown that temperate mini-Neptunes with the right properties can allow for habitable conditions in their interiors, e.g., in the case of K2-18~b \citep{madhu2020}. Therefore, in this study we focus on planets that allow for large oceans with habitable conditions underneath H$_2$-rich atmospheres. We refer to such planets as ``Hycean" worlds. 

While the potential for habitability and biosignatures of rocky exoplanets and water worlds has been studied in great detail for different atmospheric compositions as discussed above, the same has not been pursued for Hycean planets. Here, we explore Hycean planets with water mass fractions as large as 90\%, equilibrium temperatures ($T_{\rm eq}$) as high as $\sim$500~K and H$_2$-rich atmospheres as deep as 1000 bar, in search of habitable conditions. We consider `habitable conditions' at the oceanic surface to mean thermodynamic conditions known to be habitable in Earth's oceans, i.e., up to 395~K in temperature and up to $\sim$1000~bar in pressure \citep{rothschild2001,merino2019}. We explore the region in the mass-radius plane occupied by Hycean planets, and identify a sizable sample of candidate Hycean planets that are promising for atmospheric characterization. We also construct the Hycean HZ as a function of stellar type, considering a wide range of irradiation conditions including planet-wide habitability, as well as habitability of tidally locked and nonirradiated Hycean planets. Finally, we investigate the spectral signatures of several possible biomarkers and their detectability in Hycean atmospheres with transit spectroscopy. We show that Hycean planets present a new opportunity in the search for life elsewhere. 

In what follows, we investigate the bulk properties, habitability, and potential biosignatures of Hycean planets. In section~\ref{sec:mr_plane}, we first explore the region in the mass-radius plane occupied by Hycean planets and identify known planets in this regime. We then investigate, in section~\ref{sec:hab}, the atmospheric temperature structures of Hycean planets orbiting host stars across the spectral range to assess their habitability under diverse conditions. In so doing, we construct a Hycean HZ. In section~\ref{sec:biosig} we investigate the signatures and detectability of possible biomarkers in Hycean planets, using model transmission spectra of known candidates. We summarize our conclusions and discuss the implications of our results in section~\ref{sec:discussion}. 

\section{Hycean Mass-Radius Plane} 
\label{sec:mr_plane}

\begin{figure*}
\centering
\includegraphics[width=0.83\textwidth]{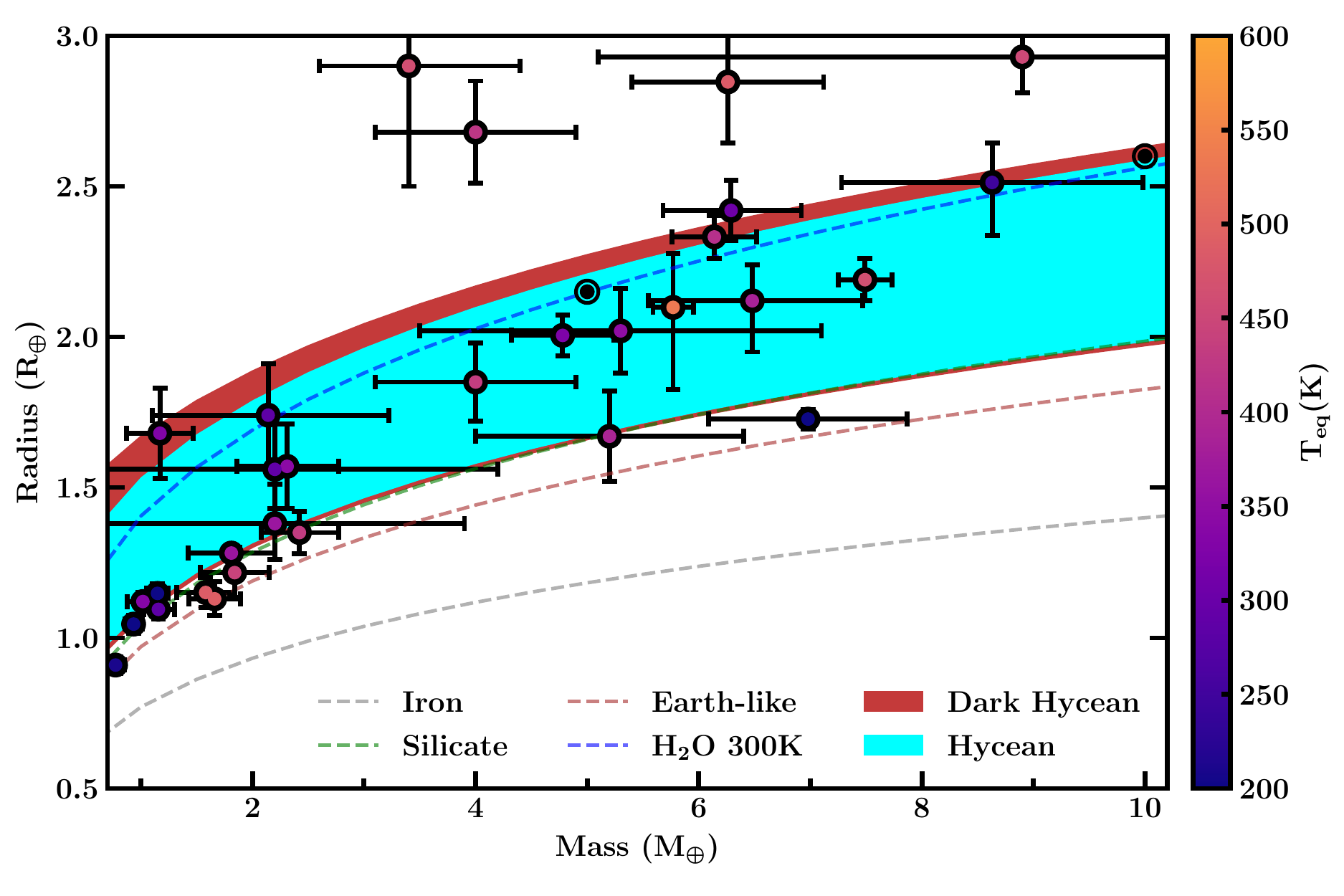}
    \caption{The Hycean mass-radius ($M$-$R$) plane. The $M$-$R$ plane of regular Hycean planets is shown in cyan, and that of Dark Hycean planets is shown in red, which includes the cyan region. Dashed lines show $M$-$R$ curves for homogeneous compositions of 100\% iron (gray), 100\% silicate (green), Earth-like composition (brown: 32.5\% Fe + 67.5\% silicate), and 100\% H$_2$O at 300~K and 1 bar surface conditions (blue), as shown in the legend. The concentric black circles show the two case studies used in Section~\ref{sec:hab}. The black circles with error bars show transiting exoplanets with observed masses and radii, color-coded by their equilibrium temperature ($T_{\rm eq}$), defined in eq~(\ref{eq:Teq}), assuming full day-night energy redistribution and a Bond albedo of 0.5. Only planets orbiting host stars with J mag $<$ 13 are shown. We note that while planets with masses and radii shown in the Hycean regions can be Hycean candidates, other internal structures may also be admissible by the data (see, e.g., Section~\ref{sec:candidates}). A list of promising Hycean candidates is shown in Table~\ref{tab:planet_list}. Exoplanet data obtained from the NASA Exoplanet Archive.}
    \label{fig:mr}
\end{figure*}

Here, we investigate the bulk properties of Hycean planets. Using internal structure models, we first identify the mass-radius plane occupied by Hycean planets and then identify known candidates for such planets. Our modeling approach closely follows that of \cite{madhu2020} on the HZ mini-Neptune K2-18~b, which is a candidate Hycean planet in the present study. 

\subsection{Internal Structure Model} 

We model the internal structure of Hycean planets following a conventional four-layered structure typically adopted for mini-Neptunes \citep[see, e.g.,][]{rogers2010,nettelmann2011,valencia2013,madhu2020}. The generic model comprises of an Fe inner core, a rocky (silicate) outer core, an H$_2$O layer, and an H$_2$/He-rich atmosphere. We refer to the Fe+silicate layers as the core. The mass fractions of each of the four components are free parameters in the model:
 $x_{\rm Fe}$, $x_{\rm silicate}$, $x_{\rm H_2O}$, and $x_{\rm H/He}$. The core mass fraction is given by $x_{\rm core}$ = $x_{\rm Fe}$ + $x_{\rm silicate}$. Given the total mass and an interior composition, the internal structure equations are solved to determine the radius of the planet. The temperature and pressure at the outer boundary are also free parameters in the model. The temperature structure in the H$_2$/He-rich atmosphere is an input to the model and sets the temperature at the H$_2$O-H$_2$/He boundary (HHB). Below the HHB the temperature structure follows an adiabatic profile in the H$_2$O layer. 

We refer the reader to \cite{madhu2012} and \cite{madhu2020} for a full description of the modeling approach. The structure equations of mass continuity and hydrostatic equilibrium are solved for the given equations of state (EOSs) in each layer using a fourth-order Runge-Kutta scheme. For the inner and outer core, we use the Birch-Murnaghan EOS \citep{birch1952} for Fe \citep{ahrens2000} and MgSiO$_3$ perovskite \citep{karki2000}, respectively, as used by \citet{seager2007}. In this work, the core is assumed to be of Earth-like composition  (32.5\% Fe and  67.5\% silicate). For the H$_2$O layer we use the temperature-dependent H$_2$O EOS adopted from  \cite{madhu2020} which is compiled from \citet{fei1993}, \citet{wagner2002}, \citet{seager2007}, \citet{french2009}, \citet{sugimura2010}; see also \citet{thomas2016} and \citet{nixon2021}. For the H$_2$/He-rich atmosphere we use the ideal gas EOS which is accurate for the low pressures and temperatures considered here given the focus on habitable ocean surfaces under the atmosphere. The mean molecular weight (MMW) of the atmosphere is set by the atmospheric composition as discussed below. 

A fiducial estimate of the possible interior composition of a planet can be obtained by considering its mass ($M_{\rm p}$) and radius ($R_{\rm p}$), i.e., its bulk density. The theoretical mass-radius ($M$-$R$) curves of planets with homogeneous compositions are shown in Figure~\ref{fig:mr} for 100\% Fe, silicate, and H$_2$O (at 300 K and 1 bar surface conditions).  We also show the $M$-$R$ curve for Earth-like composition for reference, i.e., with 32.5\% Fe and  67.5\% silicate. Given the $M_{\rm p}$ and $R_{\rm p}$ for a low-mass planet the interior composition and structure cannot be uniquely determined, as a broad range of degenerate solutions can generally explain the data \citep[e.g.,][]{rogers2010,valencia2013}. However, planets with $M_{\rm p}$ and $R_{\rm p}$ above the pure silicate curve necessitate the presence of a volatile layer (e.g., H$_2$O and/or H$_2$/He) or a mineral composition less dense than silicates, e.g., carbides \citep{madhu2012}. Furthermore, for temperate planets with $M_{\rm p}$ and $R_{\rm p}$ above the 100\% H$_2$O curve at 300 K the presence of a gaseous H$_2$/He-rich envelope becomes inevitable. As demonstrated in \cite{madhu2020}, such a scenario still allows for a degenerate set of solutions, ranging from a rocky interior with a large H$_2$-rich envelope to an ocean world with an H$_2$-rich atmosphere. In the present work we focus on the latter set of solutions, and explore the range in the $M$-$R$ plane that can be occupied by such Hycean planets. We summarize below the key model considerations made in the present work over \cite{madhu2020}, aimed specifically toward modeling Hycean planets.

Our canonical Hycean planet is composed of (a) an H$_2$-rich atmosphere, (b) an H$_2$O layer with a mass fraction between 10-90\% and a habitable surface, (c) an iron+rocky core with a minimum mass fraction of 10\%. The temperature ($T_{\rm HHB}$) and pressure ($P_{\rm HHB}$) at the HHB span $T_{\rm HHB} \sim$ 300-400 K and $P_{\rm HHB}$ = 1-1000 bar, motivated by conditions in which life is known to survive in Earth's oceans \citep{rothschild2001,merino2019}. We note that in our definition of a Hycean planet there is no landmass as the entire planet would be covered by the water layer. While the mass fraction of the H$_2$-rich atmosphere is relatively small ($\lesssim$0.1\%), it contributes significantly to the planetary radius depending on the $P_{\rm HHB}$, $T_{\rm HHB}$, gravity, and atmospheric composition.

The MMW of the  H$_2$-rich atmosphere can vary significantly depending on the atmospheric composition. For a solar abundance composition, which has an MMW of 2.4, the dominant chemical species in chemical equilibrium in the temperate regime ($\lesssim$600 K) besides H$_2$ and He are H$_2$O, CH$_4$, and NH$_3$ at volume mixing ratios $\lesssim$0.1\% \citep{burrows1999,lodders2002,madhu_seager2011}. However, for Hycean planets, with a large ocean under the atmosphere, the H$_2$O abundance in the atmosphere can be substantially enhanced. On the other hand, CH$_4$ and NH$_3$ can be depleted due to chemical disequilibrium in some conditions \citep[e.g.,][]{madhu2020, Yu2021}. H$_2$-rich atmospheres are particularly conducive to large H$_2$O enhancements \citep{koll2019}. In the present  models, we assume a 100$\times$ enhancement in the H$_2$O abundance compared to a solar abundance atmosphere, i.e., an H$_2$O volume mixing ratio of 10\% and an MMW of 4.0. This nominally includes He, CH$_4$ and NH$_3$ at abundances expected for a solar elemental composition \citep{asplund2009}. The enhanced H$_2$O mixing ratio is consistent with the upper end of the atmospheric H$_2$O abundance derived observationally for the Hycean candidate K2-18~b \citep{benneke2019,madhu2020}. 

The radii we derive with the high MMW assumed here are expected to be conservative estimates. The observable height of the atmosphere and, hence, the radius is larger for lower MMW, as the scale height is inversely proportional to the MMW. The temperature structure in the H$_2$-rich atmosphere is assumed to be isothermal in the present work, motivated by the model $P$-$T$ profiles derived in section~\ref{sec:hab}, and discussed further below. We set the outer boundary condition of the model to a pressure of 0.05 bar, following \citet{madhu2020}, corresponding to the planetary photosphere observed in transit. This is the pressure at which the radius of the planet is defined in the interior and atmosphere models  in sections~\ref{sec:hycean_mrplane} and \ref{sec:hab}, respectively.

\begin{table*}
  \caption{Properties of Promising Hycean Candidates and Their Host Stars.}
  \centering
    \begin{tabular}{l|c|c|c|c|c|c|c|c|c|c}
    \hline
    Name & $M_{\mathrm{P}}$/$M_\oplus$&  $R_{\mathrm{P}}$/$R_\oplus$ & $T_{\mathrm{eq}}$/K &  $a$/au &  $M_\star$/$M_\odot$ &  $R_\star$/$R_\odot$ &  $T_\star$/K&  $J$ mag&  $V$ mag & Ref\\
    \hline
    \hline
    K2-18~b& $ 8.63 \pm 1.35 $ & $ 2.51^{+0.13}_{-0.18} $ & 250 & 0.153 & 0.44 & 0.45 & 3590 & 9.8 & 13.5 & 1, 2\\ 
    K2-3~c& $ 2.14^{+1.08}_{-1.04} $ & $ 1.74^{+0.17}_{-0.17} $ & 286 & 0.136 & 0.55 & 0.55 & 3500 & 9.4 & 12.2 & 1, 3\\ 
    TOI-1266~c & $2.2^{+2.0}_{-1.5}$ & $1.56^{+0.15}_{-0.13}$ & 291 & 0.106 & 0.45 & 0.42 & 3600 & 9.7 & 12.9 & 4 \\
    TOI-732~c& $ 6.29^{+0.63}_{-0.61} $ & $ 2.42 \pm 0.10 $ & 305 & 0.076 & 0.38 & 0.38 & 3360& 9.0 & 13.1 & 5 \\ 
    TOI-270~d& $4.78 \pm 0.46$ & $2.01 \pm 0.07$ & 327 & 0.072 & 0.39 & 0.38 & 3506 & 9.1 & 12.6 & 6\\ 
    TOI-175~d& $ 2.31^{+0.46}_{-0.45} $ & $ 1.57 \pm 0.14 $ & 341 & 0.051 & 0.31 & 0.31 & 3412 & 7.9 & 11.7 & 7, 8\\ 
    TOI-776~c & $5.30 \pm 1.80$ & $2.02 \pm 0.14$ & 350 & 0.100 & 0.54 & 0.54 & 3709 & 8.5 & 11.5 & 9\\ 
    LTT 1445 A~b& $ 2.2^{+1.7}_{-2.1} $ & $ 1.38^{+0.13}_{-0.12} $ & 367 & 0.038 & 0.26 & 0.28 & 3337& 7.3 & 11.2 & 10\\ 
    K2-3~b& $ 6.48^{+0.99}_{-0.93} $ & $ 2.12^{+0.12}_{-0.17}  $ & 384 & 0.075 & 0.55 & 0.55 & 3500 & 9.4 & 12.2 & 1, 3\\ 
    TOI-270~c & $6.14 \pm 0.38$ & $2.33 \pm 0.07$ &  413 & 0.045 & 0.39 & 0.38 & 3506 & 9.1 & 12.6 & 6\\
    TOI-776~b & $4.00 \pm 0.90$ & $1.85 \pm 0.13$ & 434 & 0.065 & 0.54 & 0.54 & 3709 & 8.5 & 11.5 & 9 \\ 
    \hline
    
    \end{tabular}
    \begin{tablenotes}
    \item {\bf Note:} The table lists properties of promising exoplanets that fall within the Hycean boundaries in Figure~\ref{fig:mr}, with $T_\mathrm{eq} <$~500~K, and whose host stars have J~$<$~10. $T_\mathrm{eq}$ is the equilibrium temperature of the planet assuming full day-night energy redistribution and a Bond albedo of 0.5, as discussed in section \ref{sec:atmos_model}. The first five columns show the planet properties, and the following five columns show the stellar properties. $M_\star$, $R_\star$ and $T_\star$ are the mass, radius, and effective temperature of the host star, respectively.
    \newline
    \footnotesize{{\bf References:} System properties are derived from (1)~\citet{hardegree-ullman2020}, (2)~\citet{cloutier2019}, (3)~\citet{kosiarek2019}, (4)~\citet{Demory2020}, (5)~\citet{nowak2020}, (6)~\citet{VanEylen2021}, (7)~\citet{kostov2019}, (8).~\citet{cloutier2019b}, (9)~\citet{Luque2021}, (10).~\citet{winters2019}.
    }
    \end{tablenotes}
  \label{tab:planet_list}
\end{table*}

\subsection{Hycean $M$-$R$ Plane} 
\label{sec:hycean_mrplane}

We construct a Hycean $M$-$R$ plane based on the minimum and maximum radii nominally expected for Hycean planets over the 1-10 $M_\oplus$ mass range considered here. For a given mass, the factors that primarily influence the radius are $x_{\rm core}$, $x_{\rm H_2O}$, $P_{\rm HHB}$ and $T_{\rm HHB}$; the latter two parameters also influence the size of the H$_2$-rich atmosphere, which can contribute significantly to the radius. Since the core and H$_2$O layers dominate the mass content, choosing one of them naturally limits the other. Thus, the lower and upper boundaries of the Hycean $M$-$R$ plane effectively correspond to the parameter combinations that lead to the lowest and highest extent, respectively, of the H$_2$O layer and the H$_2$-rich atmosphere. 

The lower boundary of the Hycean $M$-$R$ plane is set by the minimum H$_2$O mass fraction and the minimum extent of the H$_2$-rich atmosphere possible to sustain a liquid H$_2$O ocean at the HHB. Within our Hycean model considerations this is attained for $x_{\rm H_2O}$ = 10\% and Earth-like surface conditions at the HHB, i.e., $P_{\rm HHB}$ = 1 bar and $T_{\rm HHB}$ = 300 K. From self-consistent models in section~\ref{sec:hab}, we find that the atmospheric temperature structures are nearly isothermal for most of the atmosphere. We therefore set the 1D averaged temperature profile in the H$_2$-rich atmosphere to be isothermal at 300 K for this case. Given the HHB pressure, the mass fraction of the H$_2$-rich atmosphere is $\lesssim$10$^{-5}$, while the low temperature means that the scale height is also minimal. Thus, the remaining mass is occupied by the core with $x_{\rm core}$ $\sim$ 90\%. We consider the core composition to be Earth-like, with 32.5\% Fe and 67.5\% in silicate (MgSiO$_3$) rock. 

Our consideration of a minimum $x_{\rm H_2O}$ of 10\% is such that the H$_2$O reservoir would be able to survive photodissociation and atmospheric escape over several Gyr around the most active stellar hosts \citep[e.g.,][]{luger2015b,bolmont2017}. High-energy stellar irradiation, e.g. UV activity and coronal mass ejections, can result in the photolysis of water vapor and subsequent atmospheric escape of hydrogen and oxygen. This is expected to be particularly significant for M-dwarf planets, as their host stars can be substantially more active than earlier-type stars. Through this process substantial amounts of H$_2$O can be lost, with estimates as high as 10 Earth oceans in some cases \citep{luger2015b}. We therefore conservatively set the minimum Hycean H$_2$O mass fraction to be 10\%, i.e. equivalent to $\gtrsim$100 times the Earth H$_2$O mass fraction. This amount allows the planet to retain a sizeable ocean over several Gyr even in the most active stellar environment. In practice, however, planets with even lower $x_{\rm H_2O}$ than 10\% could qualify as Hycean candidates. For example, even at $x_{\rm H_2O}$ = 1 \% a planet can be covered in oceans. In such cases, the lower boundary of the Hycean $M$-$R$ plane will be closer to the $M$-$R$ curve for Earth-like composition shown in Figure~\ref{fig:mr}. 

The upper boundary of the canonical Hycean $M$-$R$ plane is guided by the largest $x_{\rm H_2O}$ and most extended H$_2$ envelope that can still provide habitable conditions at the ocean surface. We consider this limit to be $x_{\rm H_2O}$ = 90\% and $T_{\rm HHB}$ = 400 K at $P_{\rm HHB}$ = 3 bar. This combination of $T_{\rm HHB}$ and $P_{\rm HHB}$ allows both the surface of the ocean and the atmospheric temperature to be at $\sim$400 K which is the highest temperature we consider for habitability at the ocean surface. In principle, considering a deeper HHB at $P_{\rm HHB} = 10^3$ bar at the same $T_{\rm HHB}$ can also provide a similar upper limit with a more massive atmosphere. In that case the atmospheric temperatures will be significantly cooler, leading to a smaller scale height. 

For $T_{\rm HHB}$ = 400 K the $P_{\rm HHB}$ of 3 bar ensures that the ocean surface is in liquid state; the vapor-liquid transition happens at 2.5 bar for this temperature. Above the HHB we assume the atmosphere is isothermal at 400 K, motivated by the atmospheric $P$-$T$ profiles derived in section~\ref{sec:hycean_hz}. Following \citet{madhu2020}, we consider a minimum core mass fraction of $x_{\rm core}$ = 10\% with Earth-like composition. We find that the Hycean upper $M$-$R$ boundary leads to radii that are slightly larger, by up to $\sim$0.1 $R_{\oplus}$, than those of 100\% H$_2$O planets with 300 K and 1 bar surface conditions. The radius enhancement from the higher temperature and an H$_2$-rich atmosphere is somewhat compensated by the presence of a 10\% core. We also note that assuming a lower MMW (e.g., 2.4 corresponding to 1$\times$solar metallicity) rather than the MMW of 4.0 assumed here (corresponding to an enhanced H$_2$O abundance) can lead to larger atmospheric scale heights and, hence, somewhat larger radii by up to another $\sim$0.1 $R_{\oplus}$.  

The upper boundary in the $M$-$R$ plane is even higher for partially habitable Hycean worlds. We consider the possibility of `Dark Hycean' worlds where a tidally locked planet can have a habitable permanent night (dark) side even though the permanent dayside is substantially hotter. We show in section~\ref{sec:dark_hycean_hz} that such conditions can prevail on Hycean planets with equilibrium temperatures of $\sim$510~K or even higher, depending on the dayside albedo and the day-night energy redistribution. For such planets, the temperatures at the day-night terminator as probed by transmission spectra can be significantly higher than the $\sim$400 K habitable temperature limit we consider. 

For the outer $M$-$R$ boundary of such Dark Hycean worlds, we nominally consider the planet-wide average surface and atmospheric temperature to be 500~K. The choice of this temperature is motivated by the atmospheric models for nightsides of Dark Hycean planets discussed in section~\ref{sec:dark_hycean_hz}. In particular, we find that planets with equilibrium temperatures of $\sim$510 K with inefficient day-night energy redistribution can lead to dayside temperatures of $\sim$500-600 K but nightside surface temperatures $\lesssim$400 K. Therefore, while a 510 K temperature is not considered to be habitable, it represents a planet-wide average and still allows a nonnegligible fraction of the nightside ocean surface to be at habitable surface temperatures, i.e., below 400 K. We assume a $P_{\rm HHB}$ of 30 bar which is above the 27 bar pressure required for the ocean surface to be in liquid state at the $T_{\rm HHB}$ of 500~K. The higher average temperature, compared to ``regular'' Hycean planets discussed above, both in the atmosphere and in the water layer leads to a further increase of up to $\sim$0.1 $R_\oplus$ in radius across the mass range. As shown in Figure~\ref{fig:mr}, such a condition allows for Dark Hycean worlds to be as large as  $\sim$2.6 $R_\oplus$ for $M_{\rm p}$ = 10~ $M_\oplus$.

Overall, we find that Hycean planets can occupy a wide range in the $M$-$R$ plane and can be significantly larger than super-Earths, which are assumed to be predominantly rocky. The region in the mass-radius plane occupied by Hycean worlds is shown in Figure~\ref{fig:mr}. We find that the uppermost $M$-$R$ boundary in Figure~\ref{fig:mr} allows for Dark Hycean radii that are up to $\sim$0.25 $R_\oplus$ larger than the pure H$_2$O curve at 300 K and 1 bar surface conditions; larger differences occur for lower masses. The differences are even larger compared to the $M$-$R$ curve for Earth-like composition across the mass range. Therefore, for the same mass, Hycean and/or Dark Hycean planets can be significantly larger than super-Earths and 100\% ocean worlds with habitable conditions. In the 1-10 $M_\oplus$ mass range, the upper limit on the radius of Hycean (Dark Hycean) planets is in the range of $\sim$1.5-2.6 $R_\oplus$ ($\sim$1.7-2.6 $R_\oplus$). We note that the Dark Hycean upper limit we consider here may be conservative considering that even hotter tidally locked planets than those with $T_{\rm eq}\lesssim$510~K considered here may be habitable on the nightside depending on the dayside albedo and day-night redistribution, as discussed in section~\ref{sec:dark_hycean_hz}.

On the lower boundary, Hycean radii are up to $\sim$0.2 $R_\oplus$ larger than the $M$-$R$ curve for rocky planets with Earth-like compositions, with a minimum radius of 1.1 $R_\oplus$ at $M_{\rm p}$ = 1 $M_\oplus$ and 2.0 $R_\oplus$ at $M_{\rm p}$ = 10 $M_\oplus$. Thus, between the two boundaries, Hycean planets can span a large range in masses and radii depending primarily on the mass fraction of the ocean (between 10-90\%). The H$_2$-rich atmosphere, though relatively much smaller in mass fraction ($<$ 0.1\%), can contribute significantly to the radius of the planet. Most notably, Hycean planets can be significantly larger than rocky super-Earths that are typically assumed to have Earth-like composition.  

\begin{figure*}
\centering
\begin{center}$
\begin{array}{cc}
\includegraphics[width=0.49\textwidth]{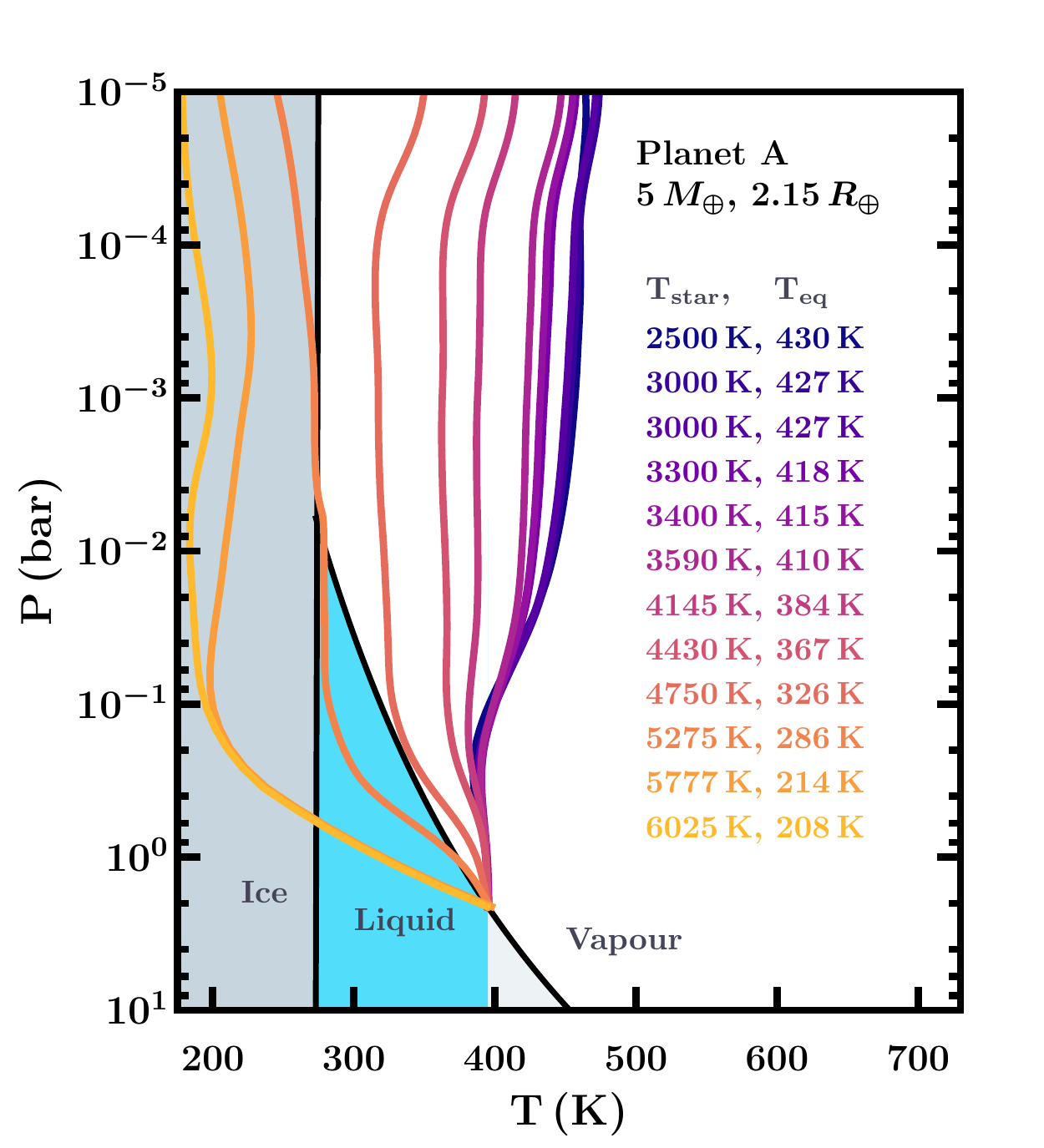}
\includegraphics[width=0.49\textwidth]{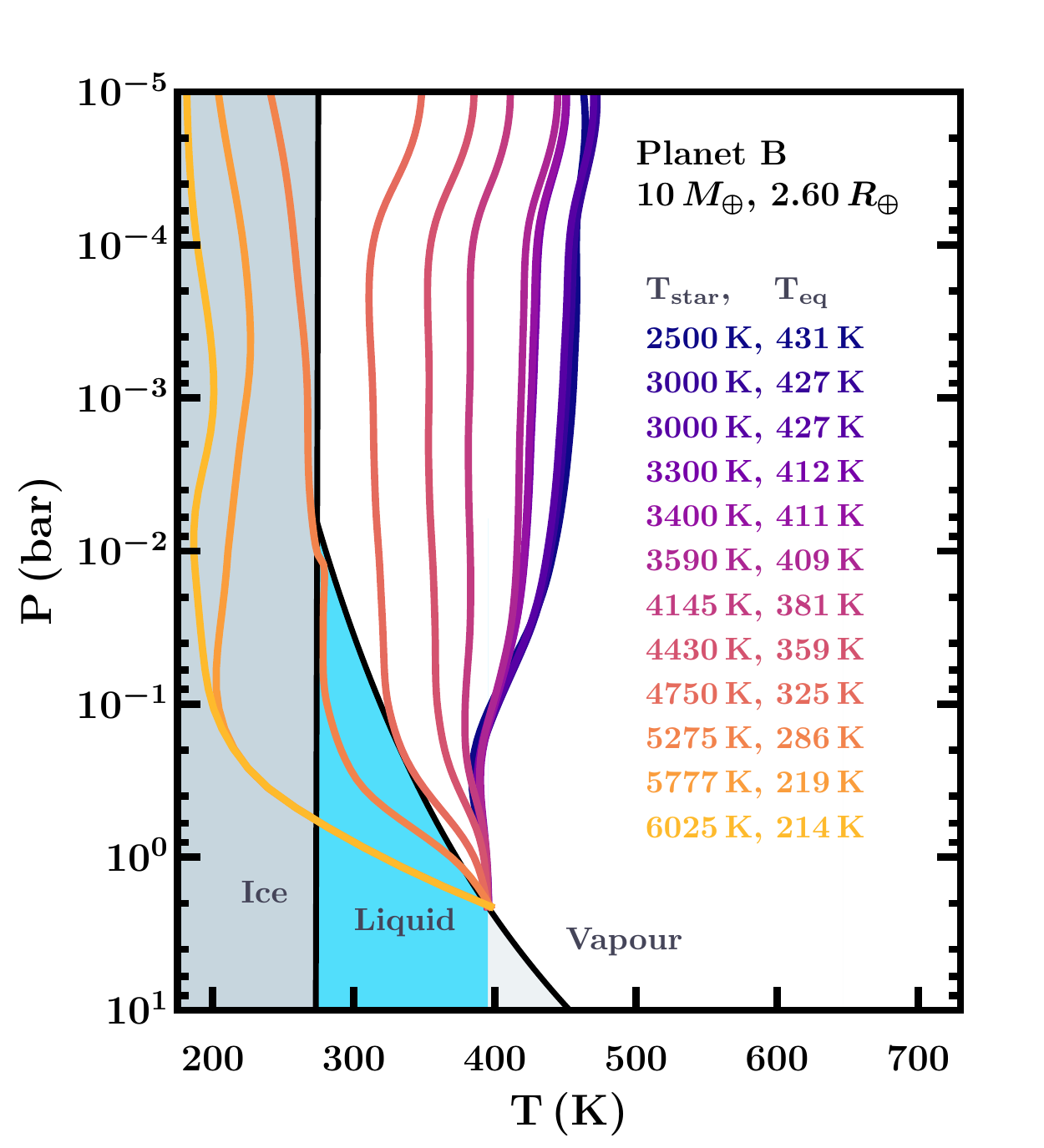}
\end{array}$
\end{center}
    \caption{Dayside temperature profiles of Hycean atmospheres with different host stars and a Bond albedo of 0.5. Left and right panels show temperature profiles for planet A and planet B, respectively. For each planet and host star combination, we find the irradiation and haze coefficient that results in a Bond albedo of 0.5 (see section \ref{sec:atmos_model}) and for which the $P$-$T$ profile reaches 395~K at 2.1~bar. The planetary equilibrium temperature, $T_\mathrm{eq}$, (for $A_{\rm B}$~=~0.5, $f_{\rm r}=0.5$) and the host star effective temperature ($T_\mathrm{star}$) are labeled in the legend. These equilibrium temperatures define the Hycean IHB (see section \ref{sec:hycean_hz}). In the background we show the phase diagram for 100\% H$_2$O, which illustrates that the phase of the H$_2$O layer beneath the atmosphere (at 2.1~bar) is liquid. The part of the liquid phase satisfying Earth-like habitable conditions (i.e. $T=$~273-395~K, $P<$~1000~bar) is highlighted in blue.}
    \label{fig:dayside_fixedAb}. 
\end{figure*}

\subsection{Hycean Candidates} 
\label{sec:candidates}

Recent transit surveys have led to numerous detections of mini-Neptunes orbiting late-type (M and K) stars \citep[e.g.,][]{dressing2015,fulton2018,hardegree-ullman2020}. Several mini-Neptunes around nearby stars are known to be conducive for atmospheric observations \citep[e.g.,][]{kreidberg2014,benneke2019,tsiaras2019,guo2020}. While the mini-Neptune class encompasses planets with radii between $\sim$1.6-4 $R_\oplus$ \citep[e.g.,][]{rogers2015}, our results above show that planets with radii between $\sim$1.1-2.6 $R_\oplus$ can be strong candidates for Hycean worlds, depending on the mass, $T_{\rm eq}$ and stellar host. In Figure~\ref{fig:mr} we show the masses and radii of several exoplanets with $M_{\rm p}<$ 10 $M_\oplus$, $R_{\rm p}<$ 3 $R_\oplus$, $T_{\rm eq} < 600$ K, and whose host stars have J magnitudes below 13. 

The identification of a Hycean candidate depends not only on its mass and radius, but also on its equilibrium temperature and stellar host. In section \ref{sec:hycean_hz}, we determine the ranges of equilibrium temperatures that allow for habitable surface conditions given a range of stellar hosts. Regular Hycean planets, with both dayside and nightside habitability, can have equilibrium temperatures as high as $\sim$210-430 K depending on the stellar host (see, e.g., Table~\ref{tab:IHB}). On the other hand, Dark Hycean planets can have planet-averaged equilibrium temperatures as high as $\sim$510~K, allowing for habitable conditions on the permanent nightside while the dayside remains uninhabitable. Therefore, all planets that lie in the Hycean $M$-$R$ plane with $T_{\rm eq} < 510$ K may be considered as candidate Hycean planets.

In Table~\ref{tab:planet_list}, we identify several Hycean candidates with masses and radii within the nominal Hycean $M$-$R$ plane that also lie within the Hycean HZ (see Figure \ref{fig:habzone}). The potentially Hycean nature of K2-18~b was demonstrated recently \citep{madhu2020}, which argues for similar conditions to be possible on the other planets listed here. Of these candidates, three lie in the Dark Hycean HZ, as shown in Figure~\ref{fig:habzone}, namely, K2-3~b, TOI-270~c and TOI-776~b. As discussed above in Section \ref{sec:hycean_mrplane}, in principle the lower Hycean boundary can be closer to the Earth-like $M$-$R$ curve if lower H$_2$O mass fractions are considered. In such a scenario, other known planets may also qualify as Hycean candidates. The planet LHS~1140~b \citep{ment2019} with $M_\mathrm{p}$ = 6.98~$M_\oplus$ , $R_\mathrm{p}$ = 1.727~$R_\oplus$ and $T_\mathrm{eq}$ = 197~K, orbiting an M4.5 dwarf star is one such potential candidate.

Figure~\ref{fig:mr} also shows a few other known planets with masses and radii in the Hycean $M$-$R$ plane but with $T_{\rm eq}>$ 510 K. While these planets would be too hot to be habitable given our current assumptions, the habitability of their permanent nightsides may not be entirely ruled out. As discussed in section~\ref{sec:dark_hycean_hz}, the allowed equilibrium temperatures of Dark Hycean worlds can be higher than 510 K for lower day-night energy redistribution efficiencies and higher albedos than those considered in this work.

We emphasize, however, that planets in the Hycean $M$-$R$ plane are only Hycean candidates. Given only the observed mass and radius of a Hycean candidate, there are significant degeneracies in establishing its internal composition and structure. A range of H$_2$, H$_2$O and core mass fractions would be admissible by the data, as demonstrated in the case of the Hycean candidate K2-18~b \citep{madhu2020}. The scenarios include internal structures ranging from mini-Neptunes and 100\% water worlds to predominantly rocky super-Earths with large H$_2$-rich envelopes. Nevertheless, accurately measured masses, radii, and equilibrium temperatures allow us to focus on promising Hycean candidates that, with spectroscopic observations, may lead to inferences of their atmospheric compositions, including biosignatures. The atmospheric properties of a Hycean candidate can provide further constraints on its surface conditions and habitability \citep[e.g.,][]{madhu2020}. 

In what follows, we model Hycean atmospheres in order to assess their habitability across a wide range of stellar spectral types. We then discuss how spectroscopic observations can constrain the atmospheric properties, including the presence of biosignatures, in Hycean planets in section \ref{sec:biosig}.

\section{Hycean Habitable Zone} 
\label{sec:hab}
We now investigate the extent of the HZ for Hycean worlds. Our goal is to assess the range of distances from a given star over which a Hycean world could maintain habitable conditions on its ocean surface, i.e. at the HHB. We explore such conditions for main-sequence stars over the late M to early G spectral types. Such studies have traditionally been conducted to establish the HZs for rocky planets with primarily terrestrial-like atmospheres dominated by heavy molecules such as N$_2$, CO$_2$, etc. \citep[]{kasting1993,selsis2007,wordsworth2013,kopparapu2013}. \citet{stevenson1999} and \citet{pierrehumbert2011} also explore the habitability of poorly irradiated rocky planets with H$_2$-rich atmospheres. Here, we explore the Hycean HZ using self-consistent model atmospheres of Hycean worlds. We consider the influence of both incident irradiation and internal flux on the dayside and nightside of irradiated Hycean planets, as well as isolated/poorly irradiated planets. 

\subsection{Atmospheric Models}
\label{sec:atmos_model}
We model the atmospheres of Hycean worlds using the \textsc{GENESIS} self-consistent atmospheric modeling framework \citep{gandhi2017,piette2020}. We consider an H$_2$-rich plane-parallel atmosphere in hydrostatic and radiative-convective equilibrium. The thermal structure is governed by radiative-convective equilibrium given the incident irradiation and internal flux, and is determined following the Rybicki scheme with complete linearization \citep{hubeny2014}. The radiation field is computed using line-by-line radiative transfer following Feautrier's method \citep{hubeny2014,hubeny2017} and the discontinuous finite element method \citep{castor1992}, as described in \citet{piette2020}. 

The temperature structure of the atmosphere depends on the external and internal energy sources, the day-night energy redistribution, and the opacity and albedo of the atmosphere. The external irradiation depends on both the host star temperature, which determines the spectral distribution of incident energy, and the total energy incident upon the planet. This total energy can be represented by the irradiation temperature: 
\begin{equation}
\label{eq:tirr}
T_{\rm irr} = \frac{T_\star}{2^{1/4}}\sqrt{\frac{R_\star}{a}}. 
\end{equation} 
$T_\star$ and $R_\star$ are the stellar effective temperature and radius, respectively, and $a$ is the orbital separation.  $T_{\rm irr}$ is equivalent to the dayside-average equilibrium temperature of the planet assuming no albedo or day-night redistribution. Correspondingly, the equilibrium temperature of the planet can be defined as
\begin{equation}
    T_{\rm eq}(A_{\rm B},f_{\rm r}) = (1-A_{\rm B})^{1/4}(1-f_{\rm r})^{1/4}T_{\rm irr},
\label{eq:Teq}
\end{equation}
where $A_{\rm B}$ is the Bond albedo and $f_{\rm r}$ is the fraction of incident irradiation redistributed to the nightside. 

In section \ref{sec:hycean_hz}, we assume uniform day-night energy redistribution (i.e., $f_{\rm r}$ = 0.5) and $A_{\rm B}$ = 0.5, representing a limiting case for determining the inner HZ boundary. Therefore, $T_{\rm eq} = 0.707\,T_{\rm irr}$. In this scenario, the equilibrium temperatures corresponding to the dayside and nightside will be equal, i.e., $T_{\rm day}$ = $T_{\rm night}$ = $T_{\rm eq}$. In section \ref{sec:dark_hycean_hz}, we investigate models with inefficient day-night energy redistribution, i.e., $f_{\rm r}<$ 0.5. In this case, $T_{\rm day}$  = $T_{\rm eq}$ but $T_{\rm night}$ =  $[f_{\rm r}/(1-f_{\rm r})]^{1/4}T_{\rm eq}$. We therefore define $T_{\rm eq,av}=T_{\rm eq}(A_{\rm B}=0.5,f_{\rm r}=0.5)$ as a representative average equilibrium temperature of the planet in this scenario.

In this work, we explore both a range of stellar temperatures (from $\sim$2500-6000~K, see below) and a range of $T_{\rm eq}$, from $\sim$0-500~K. The internal flux emanating from the planetary interior is represented by the internal temperature $T_{\rm int}$, such that the flux input at the lower boundary is given by $F_{\rm int} = \sigma T_{\rm int}^4$ \citep[see, e.g.,][]{gandhi2017}. We explore values of $T_{\rm int}$ spanning 25-50 K, as expected for sub-Neptunian planets with ages between 1-10 Gyr \citep{valencia2013}. On the nightside of a planet, day-night energy redistribution can provide a further energy source. We consider this in section \ref{sec:dark_hycean_hz} following the methods outlined in appendix \ref{sec:appendix_redist}.

The key sources of extinction in the model atmospheres are absorption from the prominent molecules and scattering from molecular H$_2$ as well as hazes. For H$_2$-rich atmospheres in the low-temperature regime, i.e. below $\sim$500~K, the prominent sources of opacity in thermochemical equilibrium are typically H$_2$O, CH$_4$, and NH$_3$ \citep[][]{burrows1999,lodders2002,madhu_seager2011,moses2013}. However, CH$_4$ and NH$_3$ can be photochemically depleted depending on the ambient conditions \citep[e.g.,][]{madhu2020, Yu2021}.
For all chemical species other than H$_2$O, which is the dominant opacity source, we nominally determine the abundances according to chemical equilibrium for the corresponding temperature structure, assuming solar elemental abundances. For a Hycean planet, H$_2$O may be expected to evaporate from the ocean surface, significantly increasing the atmospheric H$_2$O abundance compared to equilibrium values. In our models, we therefore assume a higher H$_2$O mixing ratio of 10\%, i.e., 100 times the equilibrium abundance expected for a solar-like composition, as discussed in section \ref{sec:mr_plane}. The volume mixing ratios we assume for these species are therefore $0.1$, $5.0\times10^{-4}$ and $1.3\times10^{-4}$ for H$_2$O, CH$_4$ and NH$_3$, respectively. We additionally assume a solar abundance for He. We further consider H$_2$O condensation based on the pressure-temperature ($P$-$T$) profile with respect to the H$_2$O saturation curve. In particular, we rain out any water vapor in excess of the H$_2$O vapor pressure and freeze out water vapor where it is expected to be in the ice phase.
 
We use the line-by-line opacities of these molecules computed from the corresponding line list (H$_2$O, \citealt{rothman2010} CH$_4$ and NH$_3$, \citealt{yurchenko2013, yurchenko2014} \citealt{yurchenko2011}) as well as collision-induced absorption (CIA) from H$_2$-H$_2$ and H$_2$-He \citep{richard2012}. The absorption cross sections are computed from the line lists following \citet{gandhi2017}. Besides molecular absorption, we also consider Rayleigh scattering due to H$_2$ as well as scattering from possible hazes in the atmosphere, as described below.

\begin{figure*}
\centering
\includegraphics[width=0.85\textwidth]{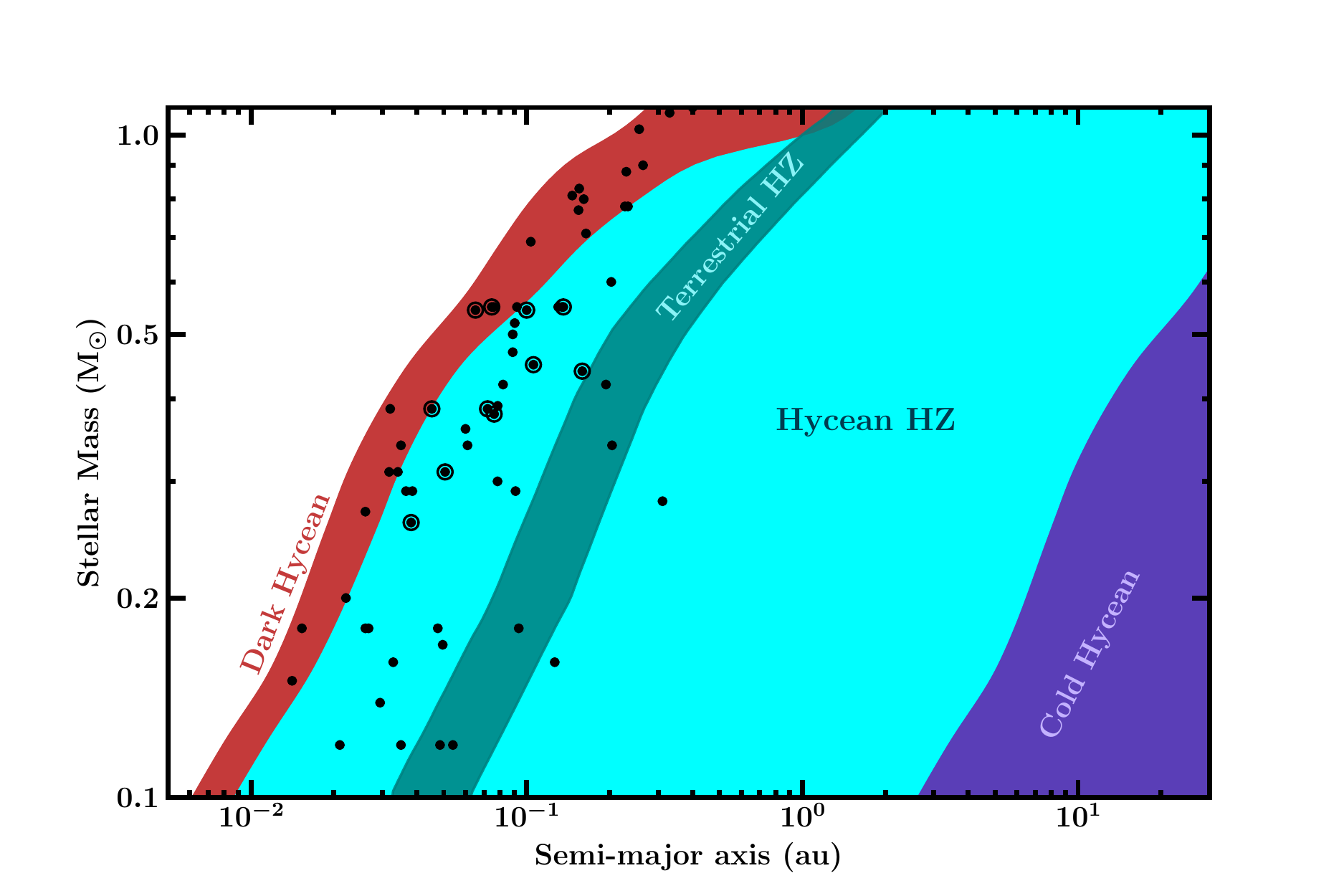}
    \caption{The Hycean HZ. Cyan, dark-red and purple regions show the HZs for regular, Dark (nightside), and Cold (nonirradiated) Hycean planets, respectively (see sections \ref{sec:hycean_hz}, \ref{sec:dark_hycean_hz} and \ref{sec:cold_hycean_hz}). The terrestrial HZ from the literature is shown in teal \citep{kopparapu2013}. Black circles denote known planets with $R_{\rm p}<$ 3 $R_{\oplus}$, $M_{\rm p}<$ 10 $M_{\oplus}$, $T_{\mathrm{eq}} <$ 600~K and whose host stars have J-band magnitudes below 13. We additionally circle the planets that are presented in Table \ref{tab:planet_list} as promising Hycean candidates. The inner edges of the Hycean and Dark Hycean HZs are calculated using planet B, which lies at the Hycean/Dark Hycean boundary in the mass-radius plane (Figure \ref{fig:mr}).} 
    \label{fig:habzone}
\end{figure*}

In order to define the Hycean HZ, we follow the approach traditionally used for determining the HZ for terrestrial-like planets \citep[]{kasting1993,selsis2007,kopparapu2013}, but tailored here for Hycean conditions. It is typical in computations of terrestrial-like HZs to assume fiducial properties for the atmospheric composition and albedo. Such computations \citep[e.g.,][]{kasting1993,kopparapu2013} account for scattering of incident irradiation by assuming a certain surface albedo without explicitly including the effect of clouds/hazes on the temperature profile. 

Here, we include the effects of hazes and parameterize their scattering as an enhanced H$_2$ Rayleigh scattering. The haze scattering cross section is given by $\sigma_{\rm haze} = {n_{\rm haze}} \sigma_{\rm H_2Rayleigh}$, where $n_{\rm haze}$ is a dimensionless free parameter and $\sigma_{\rm H_2Rayleigh}$ is the H$_2$ Rayleigh scattering cross section, following \cite{piette2020}. We refer to $n_{\rm haze}$, the Rayleigh enhancement factor, as the haze coefficient. In section \ref{sec:hycean_hz}, we calculate models with a Bond albedo of $A_\textrm{B}$~=~0.5 by varying $n_{\rm haze}$ until this target is met. This value is motivated by both previous studies \citep[e.g.,][]{selsis2007, yang2013} and the fact that most planets in the solar system have $A_\textrm{B}\sim$ 0.3-0.75 \citep{dePater2010}. We note that the majority of our models are too hot for H$_2$O clouds such as those considered in other studies \citep[e.g.,][]{morley2015,piette2020}. However, a wide range of haze compositions are thought to be possible in temperate super-Earth/mini-Neptune atmospheres \citep[e.g.,][]{moran2020}. We therefore consider haze scattering, rather than clouds, to represent the albedos across all our models for a consistent treatment.

\begin{figure*}
\centering
\begin{center}$
\begin{array}{cc}
\includegraphics[width=0.49\textwidth]{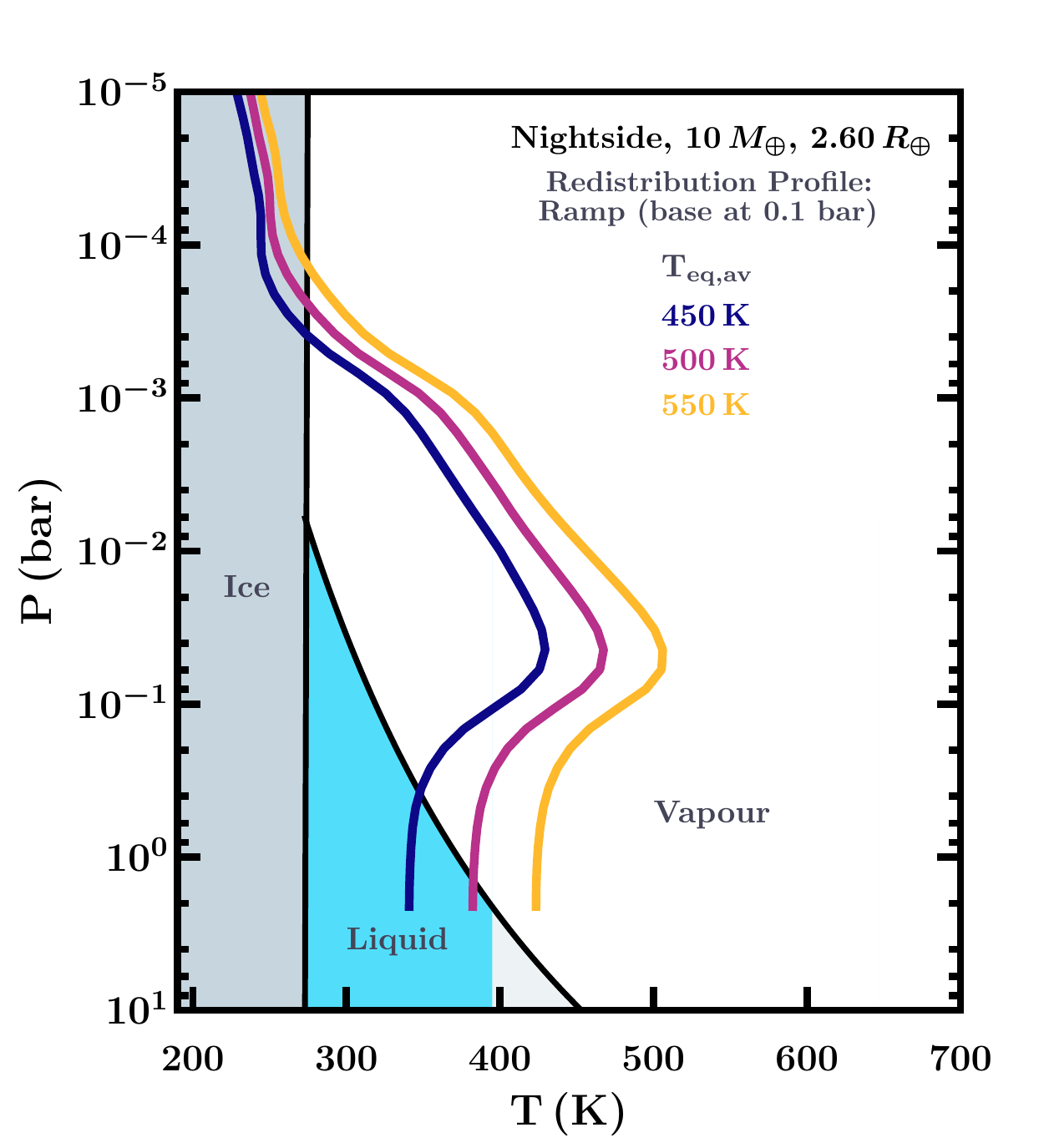}
\includegraphics[width=0.49\textwidth]{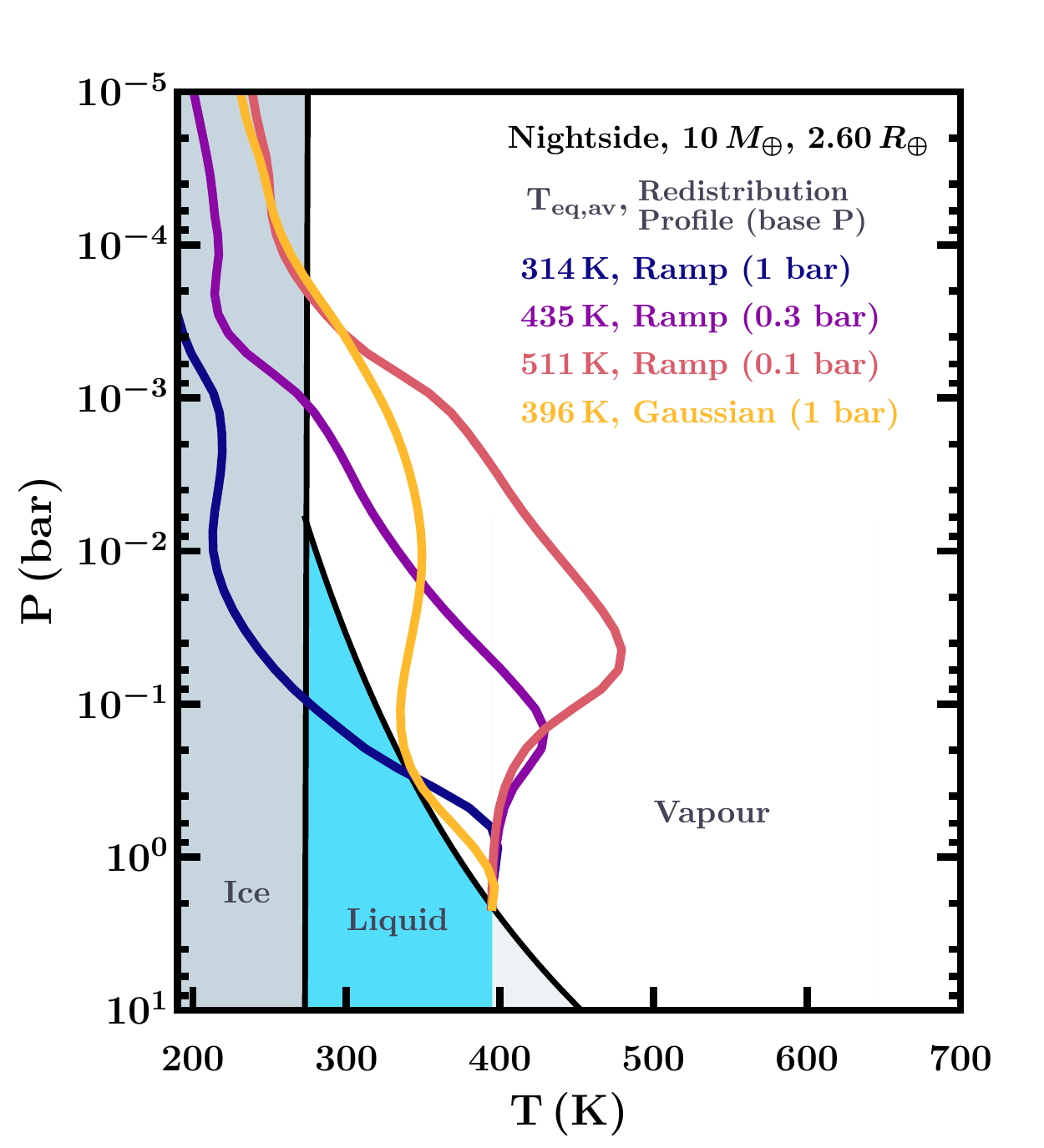}
\end{array}$
\end{center}
 \caption{Nightside temperature profiles of Dark Hycean worlds for different dayside irradiation and day-night energy redistribution profiles (see appendix \ref{sec:appendix_redist}). Left: nightside temperature profiles for different average equilibrium temperatures ($T_\mathrm{eq,av}$; see section \ref{sec:atmos_model}) and a ramp redistribution profile with a base pressure of 0.1~bar, top pressure of 1~mbar and fixed redistribution efficiency, $P_\mathrm{n}$, of 0.125 (i.e., $f_{\rm r}$~=~0.25, $A_{\rm B}=0.5$, see section \ref{sec:dark_hycean_hz} and appendix \ref{sec:appendix_redist}). Right: nightside temperature profiles for different day-night energy redistribution profiles, all assuming $P_\mathrm{n}$~=~0.125. Each redistribution profile deposits energy on the nightside at different altitudes. For each redistribution profile, we find the $T_\mathrm{eq,av}$ for which the nightside temperature profile reaches 395~K at 2.1~bar. This defines the IHB for Dark Hycean planets. Backgrounds show the phase diagram for 100\% H$_2$O, which illustrates at which temperatures the phase of the H$_2$O layer beneath the atmosphere (at 2.1~bar) would be liquid. The part of the liquid phase satisfying habitable conditions (i.e. $T=$273-395~K, $P<$ 1000~bar) is highlighted in blue.}
    \label{fig:dark_hycean_PT}. 
\end{figure*}

\begin{figure*}
\centering
\begin{center}$
\begin{array}{cc}
\includegraphics[width=0.49\textwidth]{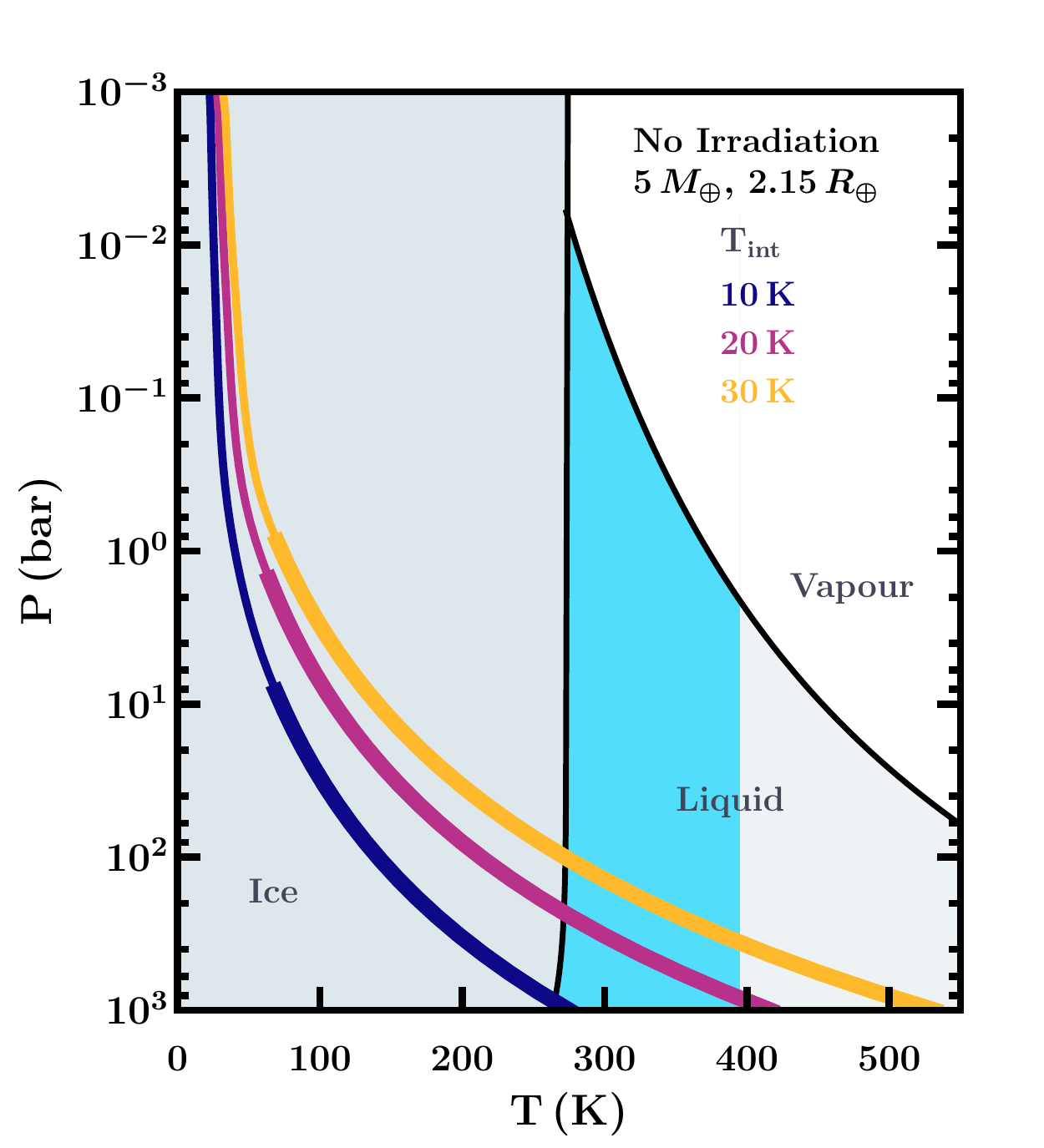}
\includegraphics[width=0.49\textwidth]{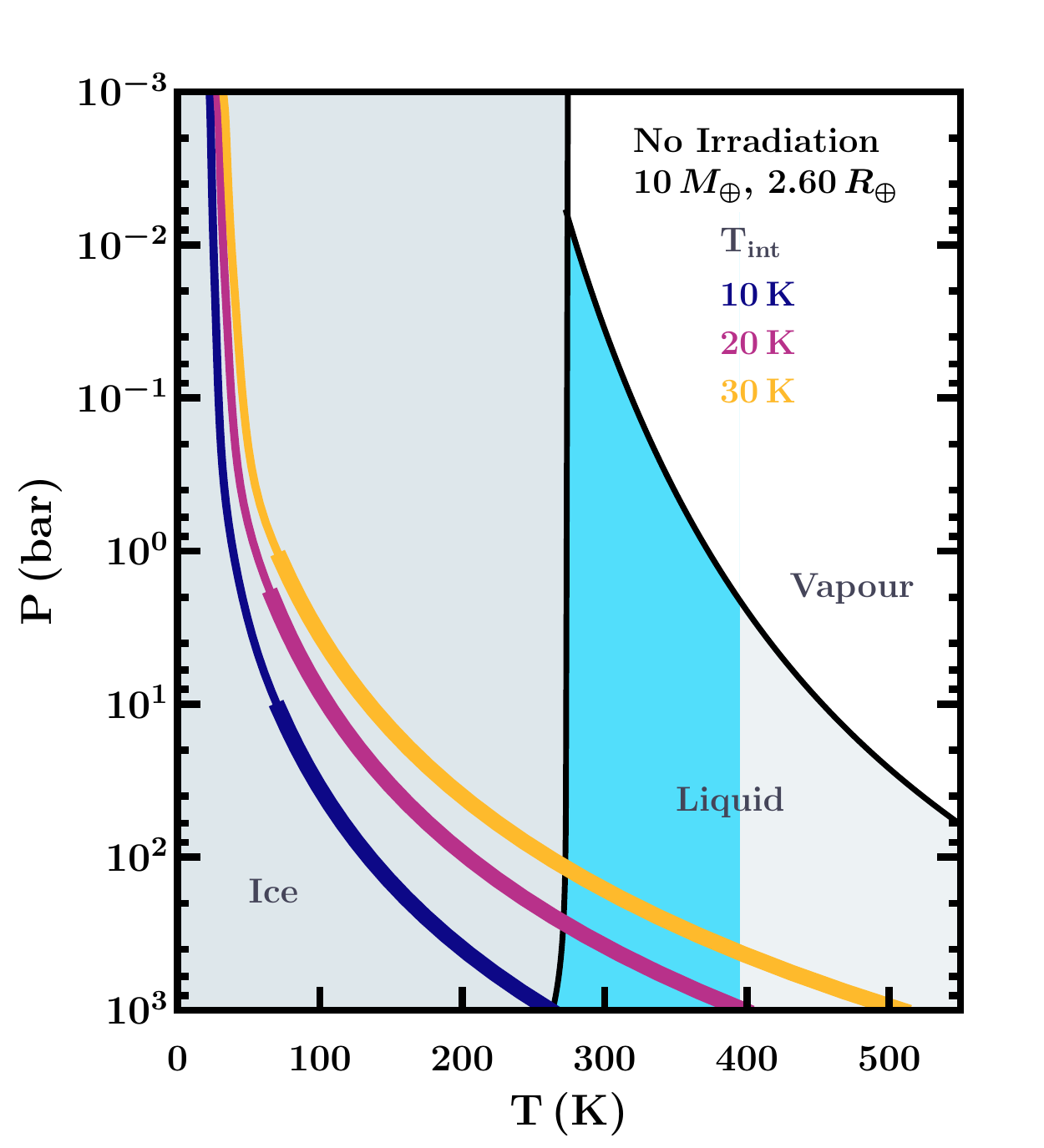}
\end{array}$
\end{center}
 \caption{Temperature profiles of Cold Hycean planets with no incident irradiation. The only energy source in these atmospheres is internal heat, characterized by $T_\mathrm{int}$. Left and right panels show temperature profiles for planet A and planet B, respectively. For each of these, we find that the lowest $T_\mathrm{int}$ that allows for habitable conditions at pressures below 1000~bar is $\sim$10~K. Thick line segments indicate convective regions in the atmosphere. In the background we show the phase diagram for 100\% H$_2$O, which corresponds to the phase of the H$_2$O layer beneath the atmosphere. The part of the liquid phase satisfying habitable conditions (i.e. $T=$~273-395~K, $P<$ 1000~bar) is highlighted in blue.}
    \label{fig:cold_hycean_PT}. 
\end{figure*}

Our conditions for habitability are motivated by the range of conditions in Earth's oceans where life is known to survive \citep[e.g.,][]{rothschild2001,merino2019}. We require that the ocean surface under the H$_2$-rich atmosphere (i.e. the HHB) is at a pressure between 1 and 1000 bar and temperature between 273 and 395~K, conditions where H$_2$O is in liquid phase and suitable for ocean-based life. This ``habitable region'' in the H$_2$O phase diagram is highlighted in blue in Figures \ref{fig:dayside_fixedAb}, \ref{fig:dark_hycean_PT} and \ref{fig:cold_hycean_PT}. We also note that H$_2$O remains in liquid form at even higher pressures and temperatures, and extraterrestrial life may acclimatize to such conditions. As such, our assumed conditions here may be considered to be conservative. We consider host stars over a wide range of spectral types spanning late M to early G, as described below. For each host star, we consider two sample planets: `planet A' with $M_{\rm p}$ = 5 $M_\oplus$, $R_{\rm p}$ = 2.15 $R_\oplus$, and `planet B' with $M_{\rm p}$ = 10 $M_\oplus$, $R_{\rm p}$ = 2.60 $R_\oplus$ (see Figure \ref{fig:mr}). For each combination of host star and planet, we consider models with a range of equilibrium temperatures. The HZ for a given star is then determined by the range in equilibrium temperature (or equivalently, orbital separation) that allows for habitable temperatures at the base of the atmosphere. 

\begin{table*}
  \centering
  \caption{Atmospheric Properties at the Inner Habitable Boundary as a Function of Host Star Properties (Effective Temperature, $T_\star$, and Stellar Mass, $M_\star$).}
    \begin{tabular}{ccccccccc}
    \hline
    \hline
          &       & \multicolumn{3}{c}{Planet A: $M_{\rm p}=5~M_\oplus$,} &       & \multicolumn{3}{c}{Planet B: $M_{\rm p}=10~M_\oplus$,} \\
          &       & \multicolumn{3}{c}{$R_{\rm p}=2.15~R_\oplus$} &       & \multicolumn{3}{c}{$R_{\rm p}=2.60~R_\oplus$}\\
\cline{3-5}\cline{7-9}    $T_\star$/K & $M_\star$/$M_\odot$ & $T_{\rm eq}$/K & $a$/au & $n_{\rm haze}$ &       & $T_{\rm eq}$/K & $a$/au & $n_{\rm haze}$ \\
    \hline
    2500  & 0.08  & 430   & 0.007 & 53000 &       & 431   & 0.006 & 67000 \\
    3000  & 0.12  & 427   & 0.011 & 31000 &       & 427   & 0.011 & 37000 \\
    3000  & 0.16  & 427   & 0.017 & 29000 &       & 427   & 0.017 & 35000 \\
    3300  & 0.26  & 418   & 0.029 & 17000 &       & 412   & 0.030 & 18000 \\
    3400  & 0.31  & 415   & 0.034 & 15000 &       & 411   & 0.035 & 16500 \\
    3590  & 0.44  & 410   & 0.057 & 12000 &       & 409   & 0.057 & 14000 \\
    4145  & 0.58  & 384   & 0.109 & 4800  &       & 381   & 0.111 & 5500 \\
    4430  & 0.69  & 367   & 0.158 & 2900  &       & 359   & 0.165 & 3100 \\
    4750  & 0.80  & 326   & 0.258 & 1200  &       & 325   & 0.261 & 1400 \\
    5275  & 0.93  & 286   & 0.487 & 450   &       & 286   & 0.487 & 520 \\
    5777  & 1.00  & 214   & 1.193 & 35    &       & 219   & 1.148 & 37 \\
    6025  & 1.18  & 208   & 1.909 & 30    &       & 214   & 1.791 & 33 \\
    \hline
    \end{tabular}%
    \begin{tablenotes}
        \item {\bf Note:} We consider a fixed Bond albedo of $A_\mathrm{B}$~=~0.5 and find the corresponding haze coefficient, $n_\mathrm{haze}$, and irradiation at the IHB (see section \ref{sec:hycean_hz}). The equilibrium temperatures quoted here assume full day-night redistribution as well as a Bond albedo of 0.5, i.e., $f_{\rm r}$~=~0.5, $A_{\rm B}$~=~0.5 in equation \ref{eq:Teq}.
    \end{tablenotes}
  \label{tab:IHB}%
\end{table*}%

For the host stars in our models, we use the properties of 12 exoplanet-hosting stars with effective temperatures, $T_\star$, in the range 2500-6000~K. This ensures that the stellar properties used are realistic and unaffected by model choices in theoretical mass-radius-temperature grids, which have previously been used in HZ studies. These stars and their properties are listed in Table \ref{tab:star_props}  (Appendix \ref{sec:star_props}). The planetary atmospheric models require two stellar inputs: the stellar radius and the stellar spectrum. In this work, we use Phoenix spectral models \citep{husser2013} for M-dwarfs with $T_\star\leq$~3500~K and Kurucz models for hotter stars \citep{kurucz1979,castelli2003}. For the Phoenix models, we round the stellar gravity, metallicity, and $T_\star$ to the nearest value in the model grid within the uncertainties. For the Kurucz models, we interpolate the spectra to the nominal stellar values. For all of the host stars, the stellar radius used is the empirical radius listed in Table \ref{tab:star_props}.

We now discuss the different scenarios that allow for habitability on Hycean planets. In what follows, the only parameters varied are $T_{\rm eq}$, $T_{\rm int}$, the haze coefficient, and the host star. All other atmospheric properties are fixed to those discussed above. 

\subsection{Hycean Habitable Zone}
\label{sec:hycean_hz}

We first investigate the HZ for the day sides of Hycean planets. Since close-in planets are largely expected to be tidally locked, this configuration is particularly relevant for transiting Hycean planets with observable atmospheres. We define the inner habitable boundary (IHB) as corresponding to the maximum irradiation that allows for habitable conditions at the surface of the ocean, i.e., the HHB. In this limit, the HHB occurs at the high-temperature/low-pressure corner of the ``habitable region'' in the H$_2$O phase diagram (defined in section \ref{sec:atmos_model}). Therefore, for these limiting cases, the $P$-$T$ profile reaches $T_{\rm HHB}$~=~395~K at a pressure of $P_{\rm HHB}$~=~2.1~bar. Since, in this scenario, the pressure at the base of the atmosphere is 2.1~bar, we compute atmospheric models up to a maximum pressure of 2.1~bar. In this limit, the H$_2$O in the atmosphere near the HHB is 10\% saturated on average across the day side. A case with 100\% saturation, which results in a similar IHB, is shown in Appendix \ref{sec:star_props}. Furthermore, in this limit, $T_{\rm int}$ has a minimal value, which we consider to be 25~K for typical Hycean worlds (see section \ref{sec:atmos_model}). 

For each host star, we determine the maximum $T_{\rm eq}$, or minimum $a$, that achieves the conditions described above. To do this, we assume a fixed albedo of 0.5 and full day-night energy redistribution ($f_{\rm r}$~=~0.5) across all stellar types, as discussed in section \ref{sec:atmos_model}. The corresponding $T_{\rm eq}$, $a$ and haze properties for each planet and host star combination are listed in Table \ref{tab:IHB}. Figure~\ref{fig:dayside_fixedAb} shows the corresponding $P$-$T$ profiles for planets A and B orbiting various stellar hosts. 

The IHB as a function of the stellar mass is shown in Figure~\ref{fig:habzone}. We find that the IHB typically occurs at smaller orbital distances relative to the terrestrial HZ, particularly for lower-mass stars. This is because the temperature profiles for planets orbiting these stars are more isothermal, allowing for hotter $T_\mathrm{eq}$ (Figure \ref{fig:dayside_fixedAb}). The isothermal temperature profiles are a result of the relatively high haze opacity, compared to solar-like stars, required to achieve an albedo of 0.5. For cooler host stars, the incident irradiation peaks in the infrared and has less flux in the optical. Therefore, to achieve the same albedo, the optical scattering in the planetary atmosphere needs to be substantially higher compared to the case for a hotter star where the irradiation peaks in the visible. 

For the coolest stars, the temperature at the base of the atmosphere can be even lower than the equilibrium temperature when $A_\mathrm{B}$ is fixed to 0.5. This is because a thermal inversion is caused at high altitudes when the high haze opacity intercepts the incident flux, similar to optical absorbers causing thermal inversions in planets orbiting hotter stars \citep{hubeny2003,fortney2008}. Therefore, for the late M host stars with $T_\star$~=~2500-3300~K, we find that the IHB for an albedo of 0.5 corresponds to $T_{\rm eq}$ $\approx$410-430~K, with an orbital separation of $\sim$0.006-0.03~au. This result is true for both planets A and B.

For hotter stars, the IHB for Hycean atmospheres approaches that of the conventional terrestrial HZ. As shown in Figure~\ref{fig:habzone}, for a sun-like star the IHB is close to that of the terrestrial case, which is expected for similar Bond albedo and host star. For the hottest host stars we consider, the temperature in the atmosphere decreases monotonically with altitude, with H$_2$O freezing out at higher altitudes and leading to a largely dry atmosphere. Overall, for the range of stellar hosts between $T_\star$ of 2500 and 6000 K that we consider, the maximum irradiation allowing habitable conditions ranges between $T_{\rm eq}$ of $\sim$210 and 430~K, corresponding to orbital separations of $\sim$0.006-1.9~au. Again, this result is similar for planets A and B.

We also investigate the outer habitable boundary (OHB) for Hycean planets. The OHB is determined by the minimum irradiation that can still allow habitable conditions at the HHB. In this limit, the HHB occurs at the low-temperature/high-pressure corner of the ``habitable region'' in the H$_2$O phase diagram (defined in section \ref{sec:atmos_model}), i.e., $T_{\rm HHB}$ = 273 K at $P_{\rm HHB}\sim$1000~bar. We find that nonirradiated, ``cold'' Hycean planets can satisfy this condition at the ocean surface. Therefore, the Hycean HZ extends to arbitrarily large orbital distances and is substantially wider than the terrestrial HZ. We discuss these `cold' Hycean planets further in section \ref{sec:cold_hycean_hz}.

In order to distinguish between irradiated and nonirradiated Hycean planets, we define a boundary between the `regular' Hycean HZ (for irradiated planets) and the `cold' Hycean HZ (for nonirradiated planets). This boundary occurs where irradiation no longer dominates the temperature profile, i.e. where $T_\mathrm{eq}\lesssim T_\mathrm{int}$ and internal heat takes over as the dominant energy source in the atmosphere. The OHB for regular Hycean planets therefore occurs at $T_\mathrm{eq}$~=~25~K, since this is the minimal value of $T_\mathrm{int}$ that we consider for Hycean planets. This OHB corresponds to an orbital distance beyond $\sim$20~au for K dwarf and G dwarf stars more massive than $\sim$0.5 $M_\odot$ and is between $\sim$2-20 au for M dwarfs. Our findings for the OHB of Hycean planets are consistent with those suggested for rocky exoplanets with H$_2$-rich atmospheres in previous studies, which focused on the low irradiation regime \citep{stevenson1999,pierrehumbert2011}. 

Overall, our results show that the HZ for Hycean planets is considerably wider than the terrestrial HZ. The Hycean IHB can be significantly closer to the host stars, i.e., with larger $T_\mathrm{eq}$, depending on the albedo. The Hycean OHB is even wider, spanning orbital distances beyond $\sim$2-100 au across the stellar types considered. In comparison, for the terrestrial HZ investigated in previous studies, the OHB is limited by CO$_2$ condensation limiting the greenhouse effect. The OHB in that case \citep{kasting1993,kopparapu2013} lies within $\sim$1.7 au for the sun and $\lesssim$0.07 au for a late M dwarf, with HZ widths of $\sim$0.7 au and $\sim$0.03 au, respectively. The terrestrial HZ can be somewhat wider depending on the model considerations \citep[e.g.,][]{selsis2007, yang2013, zsom2013} The wider Hycean HZ may increase the chances that such planets host habitable conditions, as the orbital separations required are not as restrictive compared to terrestrial planets.

\subsection{Dark Hycean HZ}
\label{sec:dark_hycean_hz}

Here we investigate the possibility of habitable conditions on the permanent nightsides of Hycean planets that are tidally locked. With no incident irradiation from the host star, the sources of energy in the nightside atmosphere of the planet are (a) energy redistributed from the dayside through atmospheric circulation and (b) the internal energy. Depending on the efficiency of day-night energy redistribution, the nightside atmosphere may allow for habitable conditions even when the dayside may not. For planets with high equilibrium temperatures and efficient day-night energy redistribution, both the dayside and nightside may be uninhabitable (e.g., for $T_{\rm eq}\gtrsim$430~K for late M host stars). However, planets with high equilibrium temperatures and inefficient day-night energy redistribution can have significant day-night temperature contrasts, and the nightsides of such planets may be habitable. Here, we define Dark Hycean planets as those that have inefficient day-night redistribution (i.e. $f_{\rm r} < 0.5$) such that only the nightside is habitable. On such planets, only nocturnal life would be possible. 

General circulation models (GCMs) of tidally locked exo-Neptunes with H$_2$-rich atmospheres show that the efficiency of day-night energy redistribution is reduced for high-metallicity atmospheres \citep{lewis2010,crossfield2020}. This can lead to significant differences in the temperatures between the dayside and nightside atmospheres. We therefore explore models with inefficient redistribution to investigate the limiting dayside irradiation that can still allow for habitability on the nightside. 

We self-consistently model the temperature structure and radiative transfer in the nightside atmosphere, accounting for energy redistributed from the dayside. Our prescription for the day-night energy redistribution is described in appendix~\ref{sec:appendix_redist}. This approach was previously developed in the context of highly irradiated hot Jupiters  \citep{burrows2008}. In the present work, we test different energy redistribution profiles, including that used in \citet{burrows2008} and a Gaussian profile, as discussed in appendix~\ref{sec:appendix_redist}. In order to determine the IHB limits across the different stellar hosts, we assume a minimum redistribution efficiency of 25\% ($f_{\rm r}$~=~0.25) as well as a dayside albedo of 0.5. Thus, of the total energy incident on the dayside, 12.5\% is redistributed to the nightside. All other atmospheric parameters are fixed to those assumed in section~\ref{sec:hycean_hz}, though for simplicity we do not include haze in the nightside models. We find that including a nominal haze opacity similar to the dayside models does not have a substantial effect on the nightside given the lack of incident irradiation. To calculate the IHB for the Dark Hycean regime, we use the bulk properties of planet B, as this lies within the Dark Hycean $M$-$R$ plane. 

The IHB for Dark Hycean planets is somewhat closer in, i.e., at higher equilibrium temperatures, than that for `regular' Hycean planets across all the host stars. As described in section \ref{sec:atmos_model}, we use a planet-wide average equilibrium temperature, $T_{\rm eq,av}$, to represent the incident irradiation. However, we highlight that the dayside and nightside of Dark Hycean planets with inefficient redistribution would have different equilibrium temperatures, allowing for a habitable nightside, while the dayside is too hot to be habitable.  Figure~\ref{fig:dark_hycean_PT} shows the nightside temperature structures of such planets with different $T_{\rm eq,av}$ and using different redistribution profiles. 

Each redistribution profile deposits energy at different altitudes on the nightside, affecting the location of the IHB. We find that habitable conditions on the nightside are possible for $T_{\rm eq,av}$ as high as $\sim$510~K. This limit occurs when the energy deposition occurs at higher altitudes  ($P\lesssim$0.1~bar). This limiting $T_{\rm eq,av}$ is independent of the stellar type since the energy redistributed to the nightside depends only on the total bolometric energy incident on the dayside. A $T_{\rm eq,av}$ of 510~K corresponds to an IHB as close as  $\sim$0.005 au for late M dwarfs and within $\sim$0.25~au for Sun-like stars, as shown in Figure~\ref{fig:habzone}. When energy is deposited at lower altitudes on the nightside , up to $P\sim$1~bar, we find that the maximal $T_{\rm eq,av}$ that allows for habitable conditions is significantly lower, at $\sim$315~K. Therefore, we consider $T_{\rm eq,av}$ of 510~K to be an upper limit for such worlds.

We note that this Dark IHB is applicable only for Hycean planets that are tidally locked. Considering the tidal-locking limit from previous studies \citep{kasting1993,selsis2007}, Dark Hycean planets may be expected to be more prevalent around low-mass stars, e.g., M dwarfs. For hotter stars the Dark Hycean IHB may be beyond the tidal-locking separation. 

The upper limit on $T_\mathrm{eq,av}$ for the Dark Hycean IHB is conservative, because for less efficient redistribution and/or higher albedo (i.e. a lower $P_\mathrm{n}$; see appendix \ref{sec:appendix_redist}) the nightside can be habitable for $T_\mathrm{eq,av}>$ 510~K. In the absence of constraints on redistribution efficiency and albedos for such planets, we nominally consider $T_\mathrm{eq,av}$~=~510~K as the upper limit. Overall, the distinguishing feature of Dark Hycean planets relative to `regular' Hyceans is that their inefficient day-night energy redistribution permits a habitable nightside while the dayside remains too hot to be habitable. On the other hand, regular Hyceans are expected to be habitable on both the dayside and nightside. The limiting planet-wide equilibrium temperature of $\sim$510 K for Dark Hycean planets is higher than that of the $\sim$430 K limit for regular Hycean planets orbiting low-mass stars. The regions in the mass-radius plane are largely similar, with the Dark Hyceans allowing for slightly larger radii, by up to $\sim$0.1 $R_\oplus$ depending on the planet mass. 

\subsection{Cold Hycean HZ}
\label{sec:cold_hycean_hz}
We also consider Hycean planets with no stellar irradiation, as would be the case for planets on very large orbital separations or for free-floating planets. We term these planets `Cold Hycean' worlds. In this scenario, the only energy source affecting the atmospheric temperature profile is internal heat. Therefore, rather than varying irradiation as in previous sections, we explore the dependence of habitability on $T_{\rm int}$. As in section \ref{sec:dark_hycean_hz}, we use the same standard composition and bulk properties for planets A and B described above and nominally do not include hazes.

We find that Cold Hycean planets can be readily conducive to ocean life. For a $T_{\rm int}$ of $\sim$10~K, the $P$-$T$ profile just reaches $\sim$270~K at 1000~bar, therefore setting the limit of the lowest $T_{\rm int}$ that allows for habitable conditions. Higher $T_{\rm int}$ then allow for habitable temperatures at shallower pressures, e.g., $T_{\rm int}$~=~30~K results in temperatures between $\sim$300-400~K at pressures between $\sim$100-300~bar. These results are true for both planets A and B, as shown in Figure~\ref{fig:cold_hycean_PT}. For planets where the HHB lies in the pressure range where $T\sim$300-400~K, a habitable ocean surface is permissible. Where the HHB is at lower pressures, the surface would be frozen but subsurface ocean life could still be possible.

As discussed in section~\ref{sec:atmos_model}, we expect the $T_{\rm int}$ of Hycean planets to lie between $\sim$25-50~K \citep{valencia2013}, thus allowing the required conditions for oceanic life on Cold Hycean planets in the far stretches of planetary systems, as well as in the interstellar medium. Our results are also consistent with those of \citet{stevenson1999} who considered thin H$_2$-rich atmospheres of rocky planetary embryos in the interstellar medium. 

\section{Biosignatures}
\label{sec:biosig}
Here we investigate the possible biosignatures of Hycean worlds and their detectability using transit spectroscopy. A Hycean world would have a fully oceanic surface with no landmass and a substantial atmosphere dominated by H$_2$, with habitable surface pressures and temperatures, as discussed in this work. Thus, any life in a Hycean world is necessarily aquatic. We do not focus on predominantly rocky super-Earths with thin H$_2$-rich atmospheres as studied previously \citep[e.g.,][]{seager2013b}. Nevertheless, any biomarkers from ocean-based life proposed in previous studies, as well as those found in H$_2$-rich conditions on Earth \citep[e.g.,][]{andrea1983, pilcher2003, segura2005,domagal-goldman2011,seager2013b,seager2016,seager2020}, may be expected to be even more prevalent in Hycean planets. In what follows, we discuss the possible atmospheric compositions and the detectability of such biomarkers in Hycean planets. 

\begin{figure*}
\centering
\includegraphics[width=0.85\textwidth]{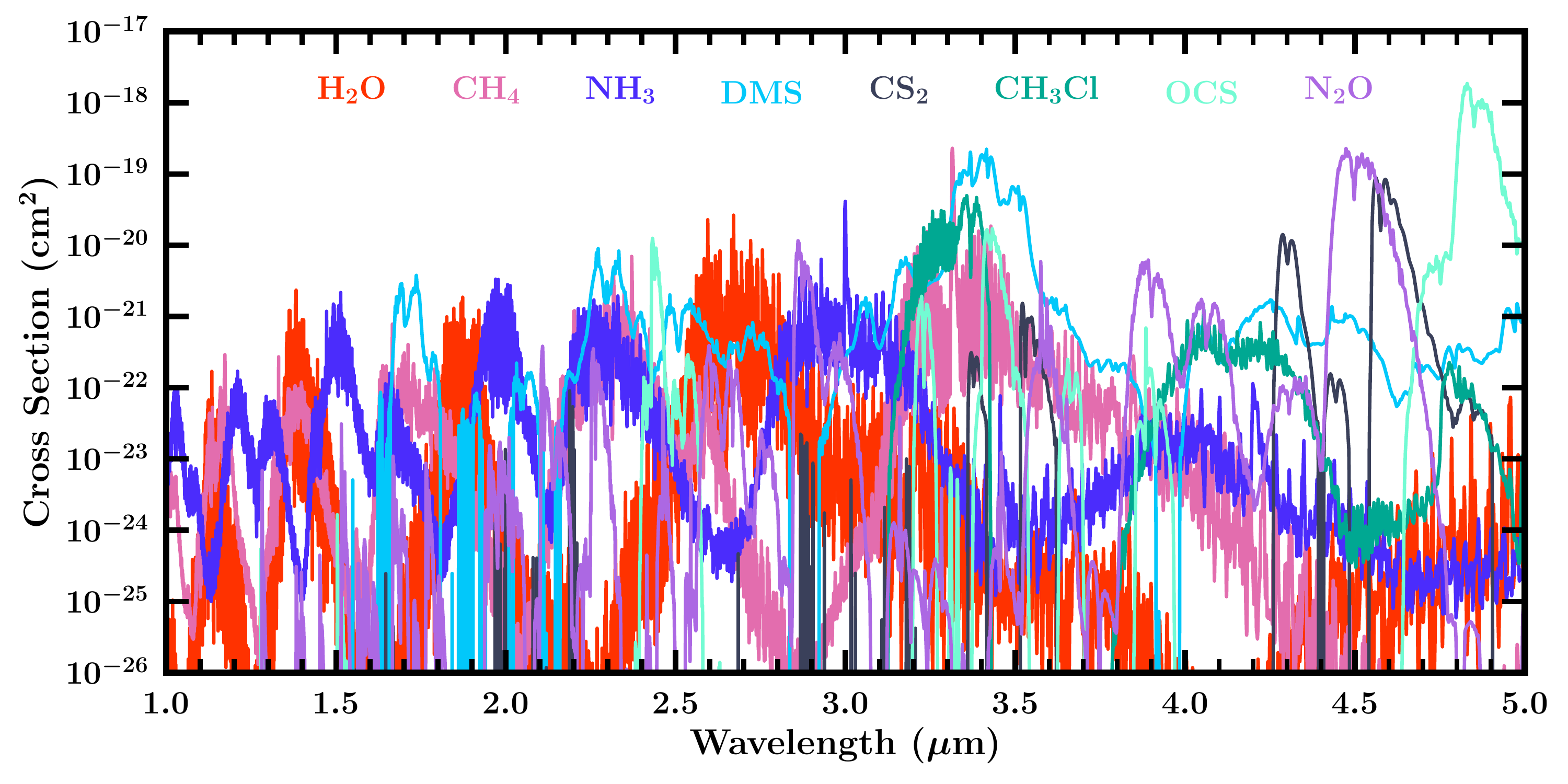}
    \caption{Absorption cross sections of key biomarkers. Cross sections are shown for the five biomarkers considered in this work (DMS, CH$_3$Cl, CS$_2$, N$_2$O, and OCS), along with other prominent molecules expected in Hycean atmospheres (H$_2$O, CH$_4$ and NH$_3$) as described in section \ref{sec:modelling_and_retrievals}. }
    \label{fig:cross_sec}
\end{figure*}

\subsection{Biosignatures in Hycean Worlds}
\label{sec:biomass}

The atmospheric composition of a Hycean planet would depend on its specific formation mechanism and atmospheric processes. Nevertheless, one may expect a general compositional framework for such a planet. Other than H$_2$/He, it is natural that H$_2$O will be a prominent constituent in such an atmosphere. As seen in solar system ice giants, CH$_4$ and NH$_3$ could also be abundant as primary  carriers of C and N, respectively \citep[e.g.,][]{atreya2016}, but they can also be depleted due to disequilibrium processes, e.g. photochemically, in Hycean conditions \citep[e.g.,][]{madhu2020, Yu2021}. All three molecules (H$_2$O, CH$_4$ and NH$_3$) can be abundant in temperate H$_2$-rich atmospheres, even assuming solar elemental ratios, and all of them have strong spectral features \citep{burrows1999,lodders2002,madhu_seager2011,moses2013}. We therefore consider H$_2$O, CH$_4$ and NH$_3$ as the dominant molecules in Hycean atmospheres providing the background opacity besides H$_2$/He, as discussed in Section \ref{sec:atmos_model}, over which signatures from any other trace species, e.g., of biomarkers, are to be detected. We also consider a case where CH$_4$ and NH$_3$ are depleted relative to equilibrium values.  

We consider five such prominent biomarkers in Hycean atmospheres: DMS, CS$_2$, CH$_3$Cl, OCS, and N$_2$O. As discussed in section~\ref{sec:intro}, these species have been suggested as potential biomarkers in atmospheres of rocky habitable exoplanets in both terrestrial-like \citep[e.g.,][]{segura2005,domagal-goldman2011,catling2018} and H$_2$-rich atmospheres \citep[e.g.,][]{seager2013b,seager2016}. \citet{seager2013b} consider a rocky super-Earth of Earth-like composition ($M_{\rm p}$ = 10 $M_\oplus$, $R_{\rm p}$ = 1.75 $R_\oplus$) with an H$_2$-rich atmosphere and estimate the abundances and detectability of these biomarkers. Their estimates suggest that all these species can be present at abundances of $\sim$1 ppmv, and up to $\sim$10 ppmv for CH$_3$Cl, and are potentially detectable in transit spectroscopy with JWST. 

Hycean atmospheres may offer even better opportunities for detecting these biomarkers than those of rocky super-earths discussed above. For a 10 $M_\oplus$ planet, the Hycean radius range is $\sim$2-2.6 $R_\oplus$ compared to the super-Earth radius of 1.75 $R_\oplus$ considered in \citet{seager2013b}. The increased radii and lower gravities lead to larger, more easily detectable spectral signatures for Hycean planets. Second, considering that prominent sources of the above biomarkers are thought to be aquatic microorganisms, we expect them to be even more abundant on Hycean worlds compared to predominantly rocky worlds. Therefore, we adopt representative abundances from \citet{seager2013b} as nominal values in our analyses below, assuming all five species to be present at 1~ppmv and allowing CH$_3$Cl abundances up to 10~ppmv, e.g., in section~\ref{sec::case_study1}. Finally, while \citet{seager2013b} advocate for NH$_3$ as a plausible biosignature gas for rocky super-Earths with H$_2$-rich atmospheres, we do not make that assumption for Hycean atmospheres where NH$_3$ can be naturally occurring as discussed above.

\subsection{Modeling and Retrieval of Transmission Spectra} 
\label{sec:modelling_and_retrievals}

We assess biosignatures of Hycean worlds that could be detectable in transmission spectra. We first investigate general characteristics of such signatures using model transmission spectra for K2-18~b, which is a candidate Hycean world. We then conduct Bayesian atmospheric retrievals of simulated spectra to assess the detectability of the biosignatures in a statistically robust manner. We model the transmission spectra using the \textsc{AURA} forward model \citep{pinhas2018}. The model computes line-by-line radiative transfer in transmission geometry assuming a plane-parallel atmosphere in hydrostatic equilibrium. The temperature structure and chemical composition are free parameters in the model. The photosphere probed by transmission spectra is typically in the 0.1-100 mbar range \citep{welbanks2019a}. As seen in section~\ref{sec:hab}, the temperature structure in the observable Hycean atmosphere is expected to be nearly isothermal in the 200-400 K range. 

We calculate the line-by-line opacities of the key molecules (H$_2$O, CH$_4$, NH$_3$), as well as H$_2$-H$_2$ and H$_2$-He CIA, in the same way as described in section \ref{sec:atmos_model}. We also consider molecular absorption due to several prominent biomarker gases predicted to be possible in H$_2$-rich environments as discussed above \citep[e.g.,][]{seager2013b,seager2020}. These include DMS, CS$_2$, CH$_3$Cl, OCS, and N$_2$O. The absorption cross sections of CH$_3$Cl, OCS, and N$_2$O were derived from the corresponding line lists from the \textsc{HITRAN} database \citep{HITRAN2016}, derived for CH$_3$Cl by \citet{ch3cl_1}, and \citet{ch3cl_2}; for OCS by \citet{ocs_1}, \citet{ocs_2}, \citet{ocs_3}, \citet{ocs_4}, \citet{ocs_5}, \citet{ocs_6}, and \citet{ocs_7}; and for N$_2$O by \citet{n2o_2}. For DMS and CS$_2$, we use the absorption cross sections provided directly by \textsc{HITRAN} \citep{dms_cs2_2,HITRAN2016,dms_cs2_1}; we assume the same cross sections across all pressures owing to the limited data available. The absorption cross sections for all the species considered in the models are shown in Figure~\ref{fig:cross_sec}, for $T$~=~300~K, $P$~=~0.1~bar. 

As can be seen from Figure~\ref{fig:cross_sec}, all these biomarkers provide significant opacity in the NIR. Importantly, several of these species provide significant opacity in the opacity windows of the more prominent molecules which may be expected in Hycean atmospheres, such as H$_2$O, CH$_4$ and NH$_3$ and are equally strong. This provides motivation to investigate the detectability of biomarkers in transmission spectroscopy of Hycean atmospheres. 

The atmospheric retrievals are conducted using an adaptation of the \textsc{AURA} retrieval code \citep{pinhas2018} as pursued in recent studies \citep[e.g.,][]{madhu2020}. We retrieve a total of 10 parameters: 8 corresponding to the volume mixing ratios of H$_2$O, CH$_4$, NH$_3$ and the 5 biomarker gases, 1 for the isotherm temperature, and 1 for the reference pressure at the fixed planet radius. For all the volume mixing ratios, we use priors that are uniform in log space, ranging from $10^{-12}$ up to $10^{-0.3}$, at which point the atmosphere can no longer be considered to be H$_2$-rich. For the isotherm temperature, the prior is uniform from 0 K to $T_{\rm eq}$+200~K for the planet under consideration. Lastly, the reference pressure prior we use is log-uniform from $10^{2}$ to $10^{-6}$ bar, which is the full atmospheric pressure range \textsc{AURA} considers in generating forward models.

\begin{figure*}
\centering
\includegraphics[width=0.88\textwidth]{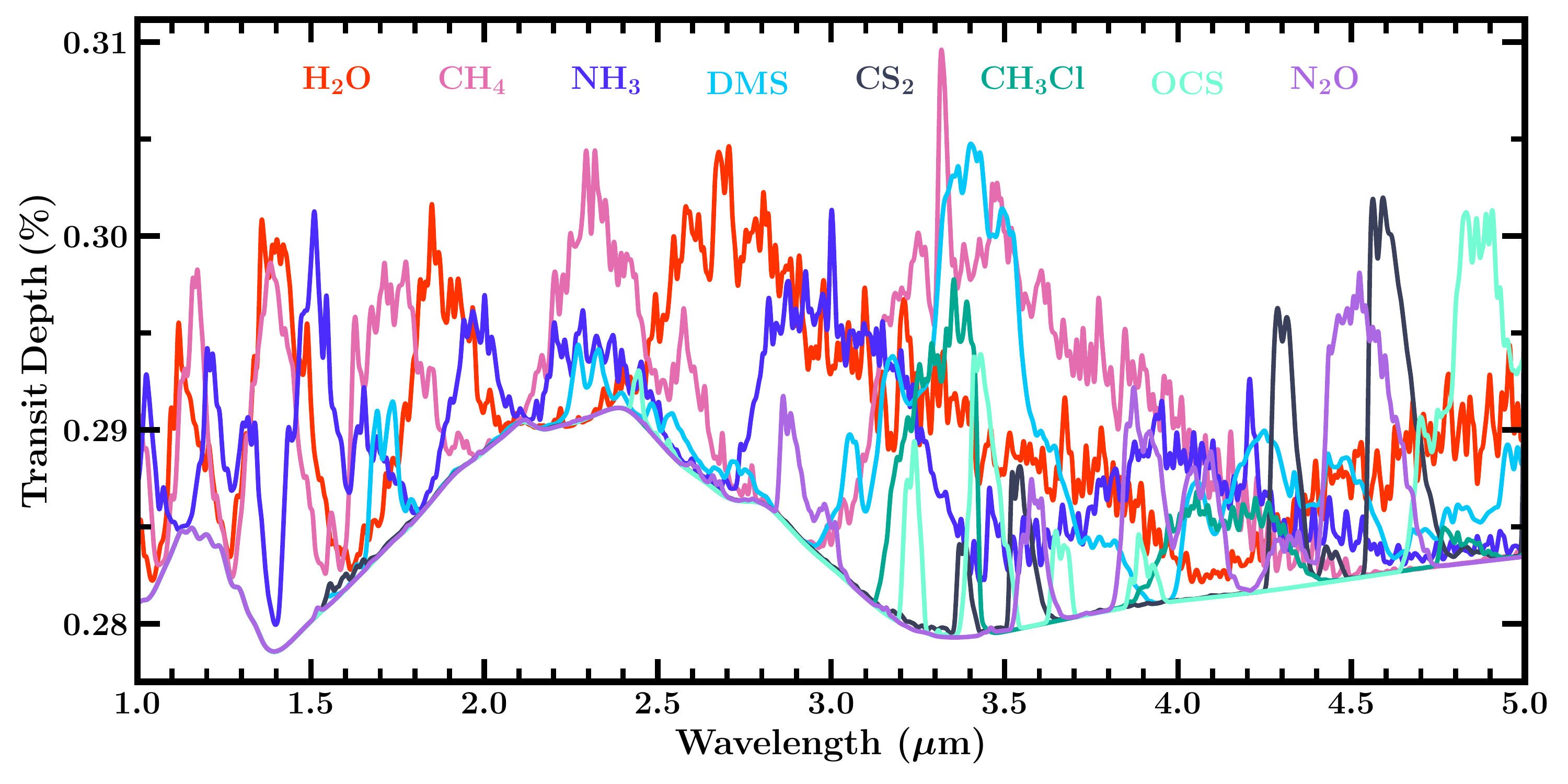}
    \caption{Molecular contributions to a model transmission spectrum of K2-18~b from the biomarkers, as well as H$_2$O, CH$_4$ and NH$_3$. Each molecule's contribution curve is the transmission spectrum generated by only including absorption from the molecule in question, as well as H$_2$-H$_2$ and H$_2$-He CIA. For each spectrum, we use the atmospheric properties and abundances for the canonical model described in section~\ref{sec::detectability}. Contributions from several biomarkers are especially prominent in the $\sim$3-5 $\mu$m range.} 
    \label{fig:contributions}
\end{figure*}

\subsection{Features in Transmission Spectra} 
\label{sec:trans_spectra}
We first assess the observable biosignatures of Hycean worlds using the exoplanet K2-18~b as a prototype. K2-18~b is the first mini-Neptune demonstrated to be potentially habitable \citep{madhu2020} and hence serves as the archetypal  candidate Hycean world. K2-18~b is a transiting exoplanet \citep{foreman-mackey2015,montet2015}, with a mass of 8.67 $\pm$ 1.35 $M_\oplus$ \citep{cloutier2019}, a radius of 2.61 $\pm$ 0.08 $R_\oplus$ \citep{benneke2019}, and a detection of H$_2$O in its atmosphere \citep{benneke2019, tsiaras2019}. We note that the radius has recently been revised to 2.51$^{+0.13}_{-0.17}$ $R_\oplus$ \citep{hardegree-ullman2020}, which is still consistent with the previous value within 1$\sigma$. In our atmospheric models for K2-18~b we use here a nominal radius of 2.61 $R_\oplus$, which agrees with both estimates, to be consistent with previous retrieval studies \citep{madhu2020}. 

The internal structure and atmospheric properties of the planet allow for the Hycean conditions described in section \ref{sec:mr_plane}. As a canonical model, we adopt representative atmospheric properties of the planet derived by \citet{madhu2020} to investigate the detectability of biosignatures in the planet's transmission spectrum. In particular, we adopt the H$_2$O abundance of 10$\times$solar, corresponding to a mixing ratio of 10$^{-2}$, which is close to the median retrieved value for K2-18~b. For CH$_4$ and NH$_3$, which were undetected in their study, we nominally assume a chemical equilibrium composition at solar elemental abundances \citep{asplund2009}, with mixing ratios of 5$\times$10$^{-4}$ and $10^{-4}$, respectively. For each of the five biomarkers we use a mixing ratio of 1  ppmv, i.e., $10^{-6}$. We assume an isothermal temperature structure at 300 K for the day-night terminator region of the atmosphere probed by transmission spectra, and we assume no clouds in the observed region, consistent with the findings of \citet{madhu2020}.

\begin{figure*}
\centering
\includegraphics[width=\textwidth]{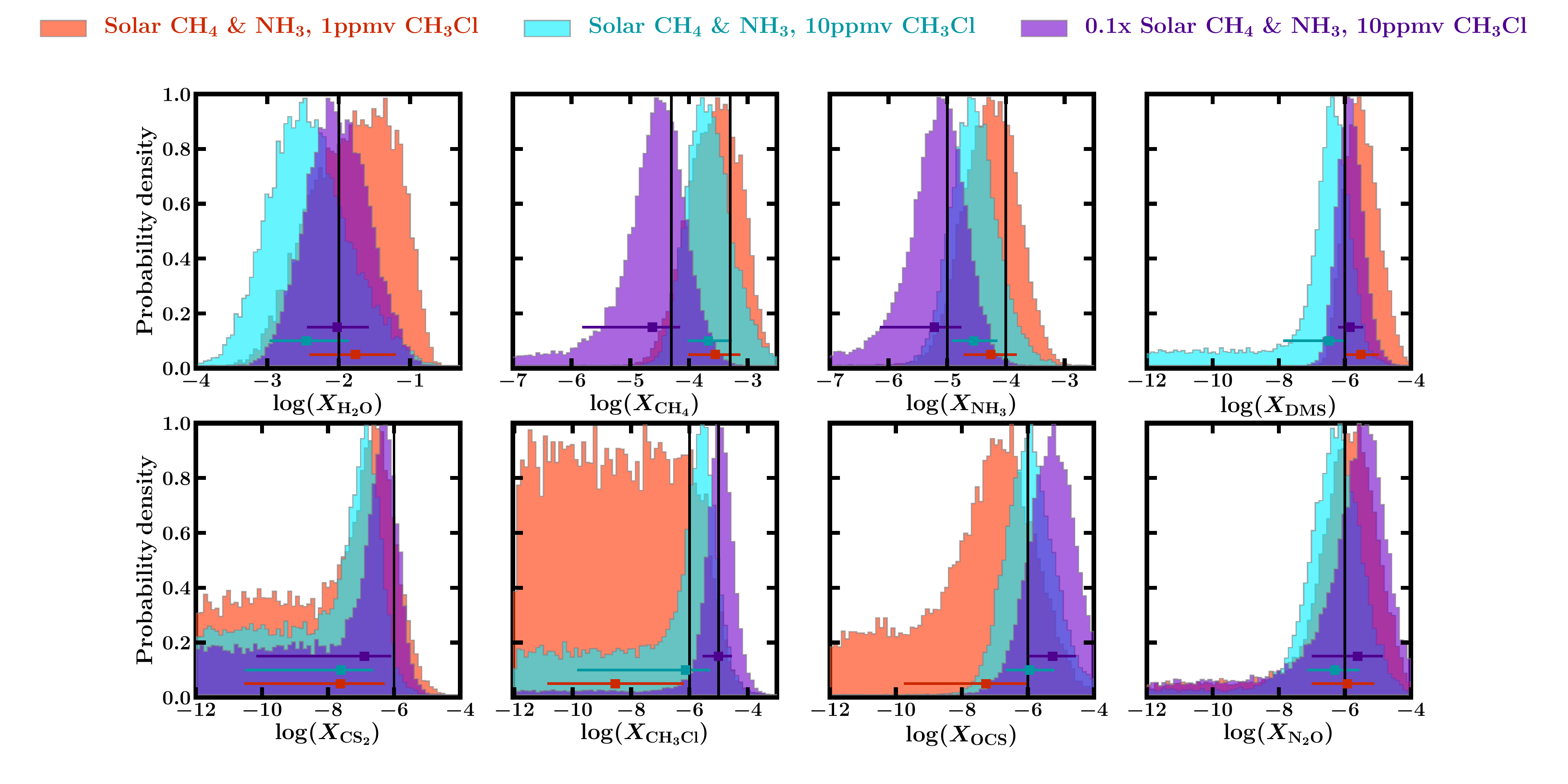}
    \caption{Posterior distributions retrieved for the mixing ratios of H$_2$O, CH$_4$, NH$_3$ and the five biomarkers from synthetic JWST transmission spectra of K2-18~b (see section~\ref{sec::detectability}). Black vertical lines denote the true input values used to generate synthetic spectra for three different cases: (i) solar abundances of CH$_4$ and NH$_3$, 1~ppmv of CH$_3$Cl (orange); (ii) solar abundances of CH$_4$ and NH$_3$, 10~ppmv of CH$_3$Cl (cyan); (iii) 0.1$\times$solar abundances of CH$_4$ and NH$_3$, 10~ppmv of CH$_3$Cl (purple). In all cases, H$_2$O is included with 10$\times$solar abundance, and the other four biomarkers have mixing ratios of 1 ppmv. Median retrieved values and 1$\sigma$ intervals are shown by the colored squares and corresponding error bars.} 
    \label{fig:K2-18b_multi}
\end{figure*}

We find strong contributions from all five potential biomarkers in the NIR region, particularly between 1.5-5 $\mu$m, as shown in Figure~\ref{fig:contributions}. The strongest contributions in this spectral range are seen for DMS, with multiple strong absorption peaks, especially at 3.4 $\mu$m and 4.2 $\mu$m, where there are not many other significant peaks. While the expected abundance of DMS is lower compared to that of the prominent molecules (H$_2$O, CH$_4$ and NH$_3$), its strong absorption cross section in this wavelength range makes it readily detectable in transmission spectra. CS$_2$ and CH$_3$Cl also have a few comparable peaks in absorption in the 3-5 $\mu$m range. We note that most of the significant peaks from CH$_3$Cl coincide with regions where CH$_4$ also has large features. OCS and N$_2$O also have significant contributions to the spectrum, which enables their detectability. Their prominent absorption peaks, however, are over a narrower wavelength range, between 4-5 $\mu$m. We also note the strong contribution from CIA opacity in the $\sim$2-3 $\mu$m range that provides strong continuum opacity, capable of masking line absorption from some of the molecules considered within that range. Furthermore, the absorption cross sections of the biomarkers used here are somewhat limited to terrestrial conditions. More extensive absorption data in the future for Hycean conditions may refine the detectability estimates in this study.

\begin{figure*}
\centering
\includegraphics[width=0.84\textwidth]{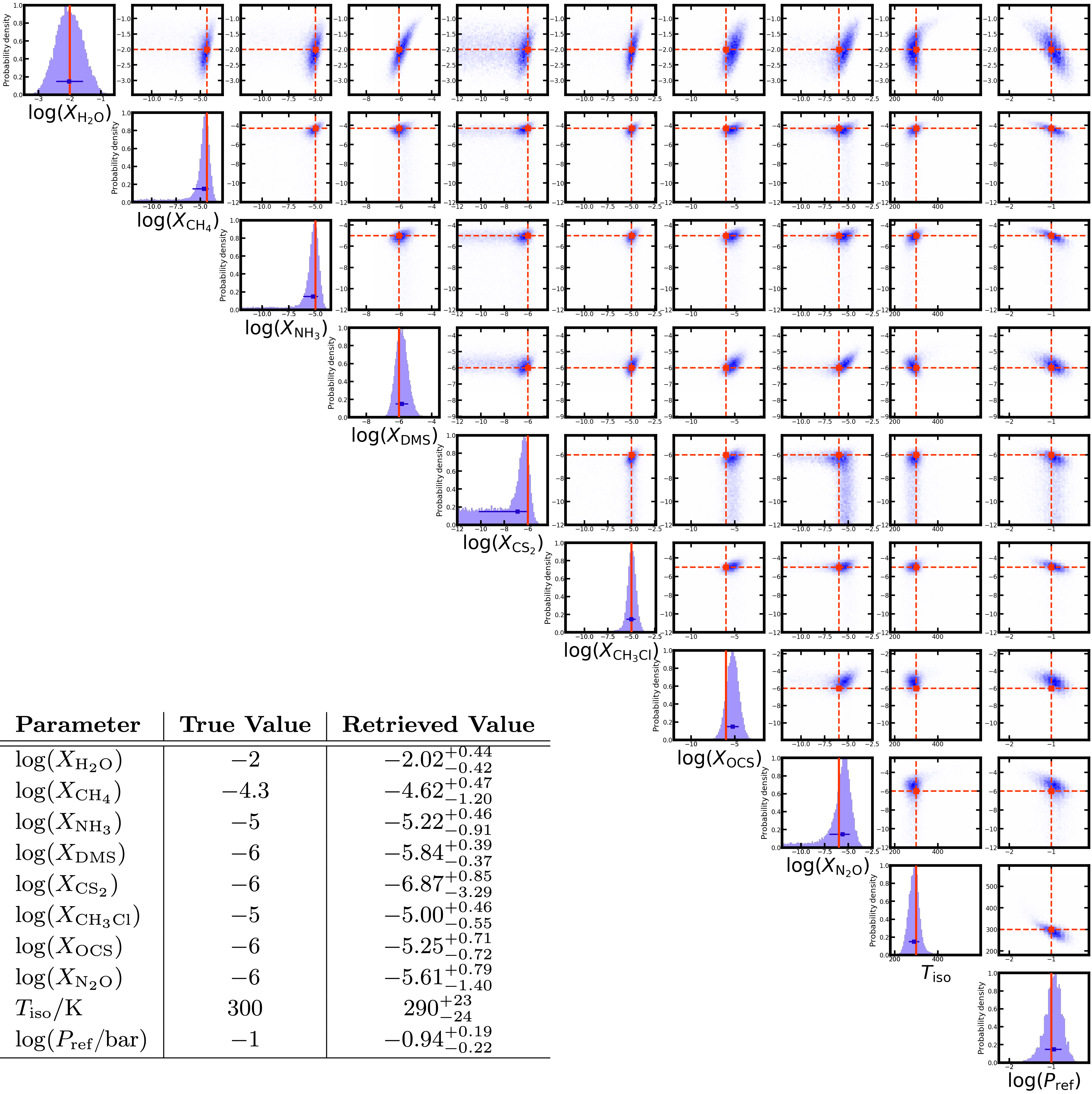}
    \caption{Marginalized posterior probability distributions from the retrieval of a synthetic transmission spectrum of K2-18~b. The parameters include mixing ratios for eight molecular species (including five biomarkers), the isothermal atmospheric temperature ($T_\mathrm{iso}$), and the reference pressure ($P_\mathrm{ref}$) where the planet radius is defined. This corresponds to the case with an H$_2$O abundance of 10$\times$solar, CH$_4$ and NH$_3$ abundances of 0.1$\times$solar, CH$_3$Cl at 10 ppmv and all other biomarkers at 1 ppmv abundances (see section~\ref{sec::case_study1}). Input parameters for the synthetic spectrum are shown by vertical red lines for the 1D distributions and by the dashed red lines and squares for the correlation plots. Median retrieved values and 1$\sigma$ intervals are shown by the dark-blue squares and error bars in the 1D posterior distributions. The true and retrieved values are listed in the table for each parameter.}
    \label{fig:K2-18b_corner}
\end{figure*}

\begin{figure*}
\centering
\includegraphics[width=\textwidth]{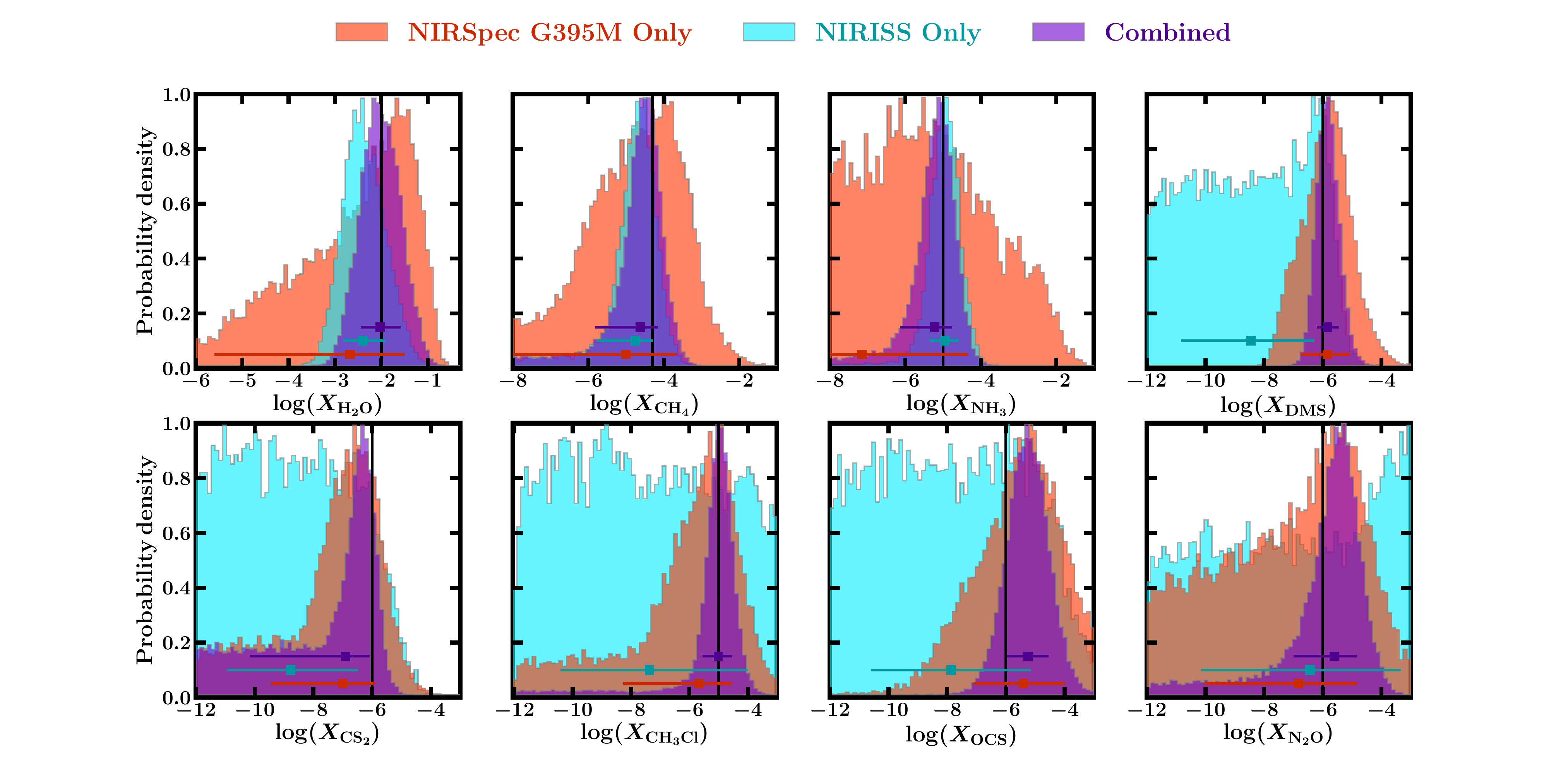}
    \caption{Posterior distributions retrieved for the mixing ratios of H$_2$O, CH$_4$, NH$_3$ and five key biomarkers from a synthetic transmission spectrum of K2-18~b for different instrument combinations. In all three cases, we use the same atmospheric parameters as shown in figure \ref{fig:K2-18b_corner}. Orange, cyan, and purple distributions correspond to synthetic spectra obtained from NIRSpec G395M only, NIRISS only, and both instruments combined, respectively. Black vertical lines denote the true input values used to generate the synthetic spectra. Median retrieved values and 1$\sigma$ intervals are shown by the colored squares and corresponding error bars.} 
    \label{fig:K2-18b_wavcov}
\end{figure*}

\begin{figure*}
\centering
\includegraphics[width=\textwidth]{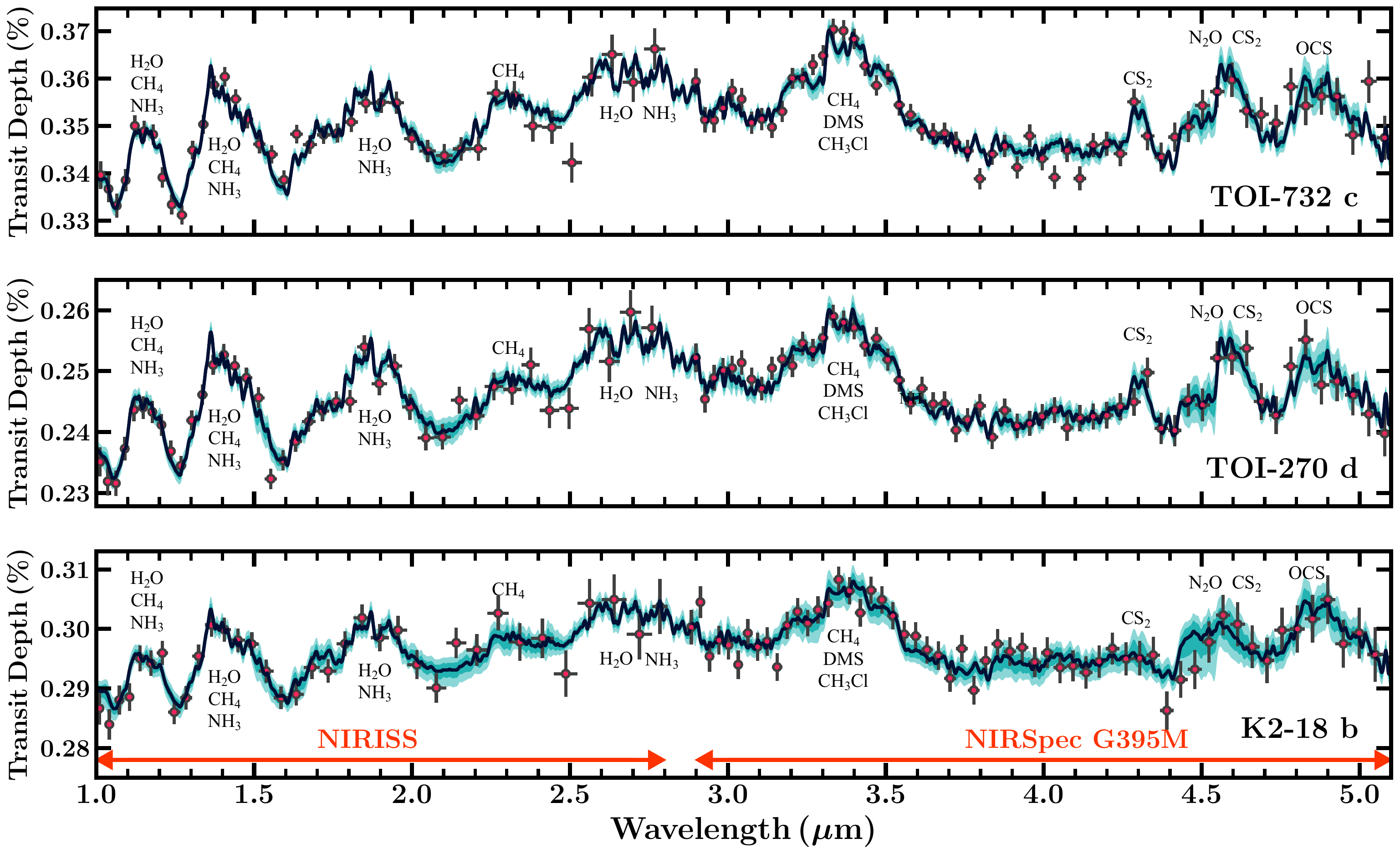}
    \caption{Retrieved spectral fits to the synthetic data for K2-18~b (bottom), TOI-270~d (middle), and TOI-732~c (top). The synthetic data for NIRISS and NIRSpec G395M are shown as circles with error bars, binned to $R$~=~40 and $R$~=~100 respectively, for visual clarity. Also shown are the corresponding median retrieved transmission spectra (dark blue lines). Darker and lighter cyan shaded regions denote the 1-~ and 2-$\sigma$ intervals for the retrieved spectrum, respectively. We additionally label the species whose opacities give rise to each significant absorption peak. The synthetic data are described in sections \ref{sec::case_study1} and \ref{sec::case_study2}.}
    \label{fig:spectra}
\end{figure*}

\subsection{Detectability of Biosignatures} 
\label{sec::detectability}
We now assess the robustness with which the biosignature molecules discussed above can be detected in Hycean atmospheres. Considering that most of the NIR spectral features of these molecules are in the 1.5-5 $\mu$m range, we consider their detectability with instruments aboard JWST that operate over this spectral range \citep[e.g.,][]{greene2016,batalha2018,kalirai2018, Sarkar2020}. Our approach here is to first generate a synthetic transmission spectrum for a planet assuming a given set of atmospheric properties. We then conduct atmospheric retrievals of the synthetic spectrum to assess which of the molecules can be confidently detected in these atmospheres and under what conditions.

For given planetary parameters, we generate a synthetic transmission spectrum in the 0.5-5.5 $\mu$m range. The canonical model spectrum assumes the molecular abundances(volume mixing ratios) given in section~\ref{sec:trans_spectra}, namely, $X_{\rm H_2O}$ = 10$^{-2}$, $X_{\rm CH_4}$ = 5$\times$10$^{-4}$, $X_{\rm NH_3}$ = 10$^{-4}$, and all five biomarkers at 1 ppmv, i.e., $10^{-6}$ each. The temperature structure is assumed to be isothermal at 300 K and the atmosphere is assumed to be cloud-free in the observable atmosphere. Beyond this canonical model, we also investigate other conditions in cases discussed below. 

We generate synthetic data using the Pandexo software package \citep{Batalha2017}, which allows for simulation of JWST observations. We provide a high-resolution forward model to Pandexo, which then yields the appropriate wavelength bins and corresponding uncertainties for the particular planet under consideration and chosen instrument settings. We then bin a high-resolution forward model to the Pandexo-provided bins, accounting for each instrument's spectral point spread function and overall transmission function. We lastly introduce noise to the synthetic data by adding an offset to each datapoint, drawn from a Gaussian distribution with standard deviation equal to the Pandexo uncertainty in that bin.

We simulate observations with NIRISS Order 1 \citep{Doyon2012} and NIRSpec G395M \citep{ferruit2012, Birkmann2014}, achieving a wavelength coverage between 1-5.1 $\mu$m. We consider a baseline configuration requiring only modest observing time with JWST: one transit with NIRISS and three transits with NIRSpec. For NIRISS we simulate one observed transit using the GR700XD grism, subarray SUBSTRIP96, and the NISRAPID readout mode. For NIRSpec G395M, we simulate three observed transits using the F290LP filter, NRSRAPID readout mode, and the SUB2048 subarray for maximal wavelength coverage. Binned to $R$ = 100, our simulated NIRISS and NIRSpec G395M observations have average uncertainties of $\sim$40 and $\sim$30 ppm, respectively, for the case of K2-18~b. Similar uncertainties at lower resolution can be achieved in the NIR (1.1-1.7 $\mu$m) with the HST WFC3 spectrograph for super-Earths and mini-Neptunes \citep[e.g.,][]{kreidberg2014,benneke2019,guo2020}. We note that the amount of JWST observing time needed for such observations corresponds to a Small or Medium General Observer Proposal, depending on specific system parameters and overheads, while even more precise observations than these are possible by dedicating more JWST observing time.

\subsubsection{Case Study: K2-18~b}
\label{sec::case_study1}

We first consider the case of the Hycean candidate planet K2-18~b. We explore the detectability of biomarkers under different assumptions for their abundances relative to those of the dominant molecules in the atmosphere. We start with a synthetic model spectrum based on the canonical abundances described above, i.e., the dominant molecules at $X_{\rm H_2O}$ = 10$^{-2}$, $X_{\rm CH_4}$ = 5$\times$10$^{-4}$, $X_{\rm NH_3}$ = 10$^{-4}$, and all five biomarkers at 1 ppmv, i.e., $10^{-6}$ each. We then investigate deviations from this canonical model and its effect on the detectability of the biomarkers. In each case, we create synthetic data based on the assumed model composition as described above and then retrieve it to assess the accuracy and precision with which the biomarkers can be retrieved. The retrieved posterior distributions for three different compositions are shown in Figure~\ref{fig:K2-18b_multi}. We find that for the canonical model the dominant molecules H$_2$O, CH$_4$, and NH$_3$ are retrieved accurately, with the true values lying within the 1$\sigma$ uncertainties of $\sim$0.6 dex for H$_2$O and $\sim$0.5 dex for CH$_4$ and NH$_3$. Additionally, two of the five biomarkers, DMS and N$_2$O, are also retrieved accurately at their trace values of 1~ppmv with uncertainties of $\sim$0.5 and $\sim$0.9 dex, respectively. Two more biomarkers, CS$_2$ and OCS, also have posterior distributions showing significant peaks near the correct mixing ratios, but with larger uncertainties. However, we do not constrain CH$_3$Cl at this abundance, instead finding only an upper limit (99\% confidence) of $\sim$10$^{-5}$. 

\begin{figure*}
\centering
\includegraphics[width=\textwidth]{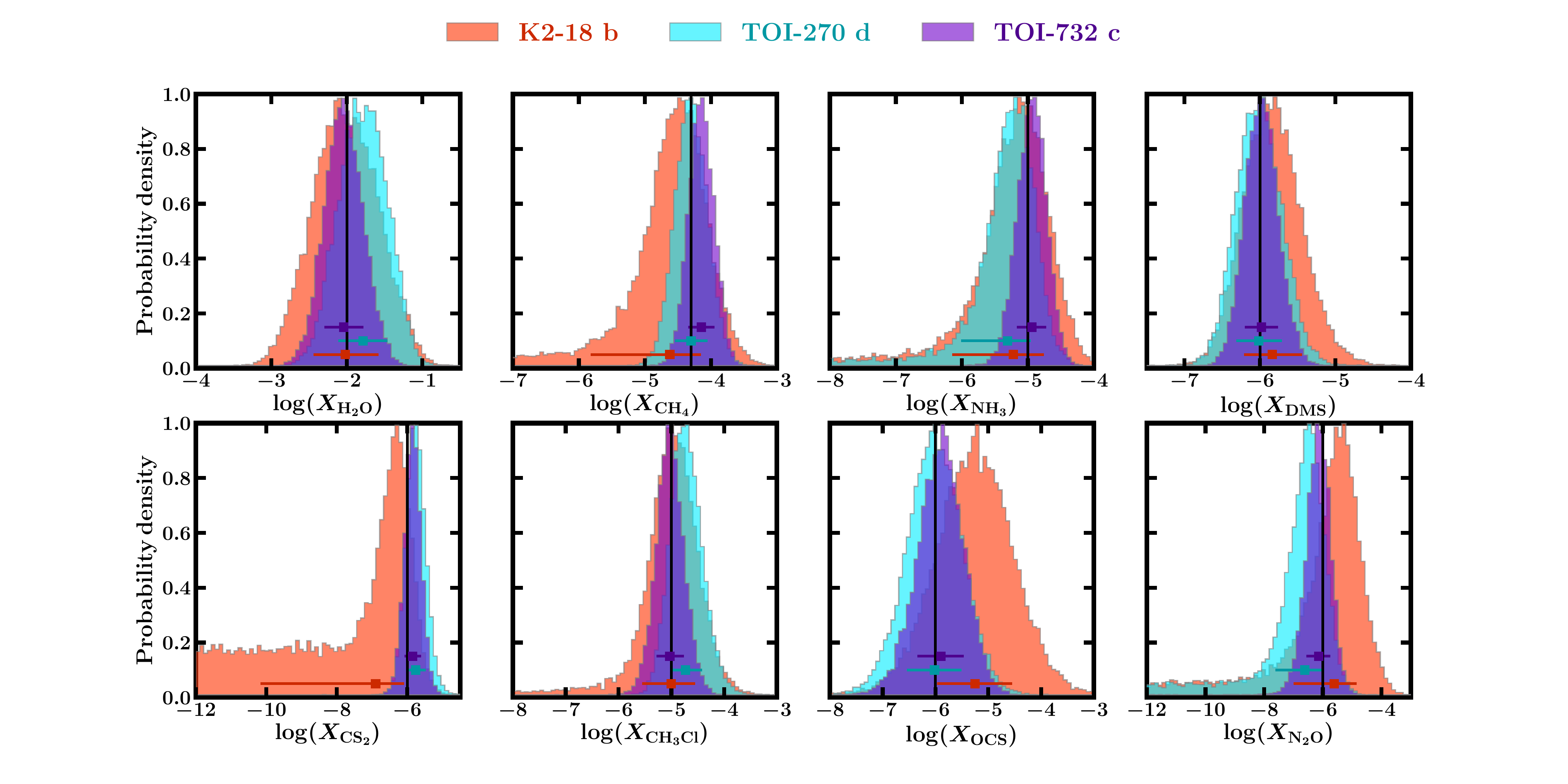}
    \caption{Posterior distributions retrieved for the mixing ratios of H$_2$O, CH$_4$, NH$_3$ and the five biomarkers from synthetic transmission spectra of K2-18~b (orange), TOI-270~d (cyan), and TOI-732~c (purple) (see sections \ref{sec::case_study1} and \ref{sec::case_study2}). Black vertical lines denote the true input values used to generate the synthetic spectra. Median retrieved values and 1$\sigma$ intervals are shown by the colored squares and corresponding error bars. We use the same atmospheric composition as used in figure \ref{fig:K2-18b_corner}.}
    \label{fig:planet_comp}
\end{figure*}

The nondetection of CH$_3$Cl at 1 ppmv is due to the fact that its strongest absorption feature, lying between 3-3.5 $\mu$m, is masked by stronger absorption at the same wavelengths by the more abundant CH$_4$ as well as equally abundant DMS, besides minor contributions from other species, as seen in Figure~\ref{fig:contributions}. Similarly, its absorption peak between 4-4.5 $\mu$m also overlaps with stronger contributions from other molecules. However, we are able to better constrain CH$_3$Cl if in the synthetic model we either (a) increase its abundance by 1 dex (to 10 ppmv) or (b) decrease the abundance of CH$_4$ by 1 dex to 0.1 $\times$solar or  5$\times$10$^{-5}$. Both these scenarios are plausible in K2-18~b; the lower CH$_4$ abundance is consistent with its nondetection in previous studies \citep{benneke2019,tsiaras2019,madhu2020}, and the 10 ppmv CH$_3$Cl abundance is plausible based on the biomass estimates of \citet{seager2013b}, discussed in section~\ref{sec:biomass}. While reducing the abundance of CH$_4$ alone is enough to constrain  CH$_3$Cl, previous atmospheric retrievals of K2-18~b have also resulted in nondetections of NH$_3$. In subsequent retrievals, we therefore vary the abundances of CH$_4$ and NH$_3$ together to maintain the solar C/N ratio and also because both of them can be depleted due to disequilibrium processes. In the case where the abundance of CH$_3$Cl is increased to 10~ppmv it becomes better constrained, albeit still with a large uncertainty, as shown in Figure~\ref{fig:K2-18b_multi}. Similar results are obtained when the abundances of CH$_4$ and NH$_3$ are instead decreased to 0.1$\times$solar. 

The last scenario we consider involves both decreasing the abundances of CH$_4$ and NH$_3$ to 0.1 $\times$solar and also increasing the CH$_3$Cl abundance to 10ppmv. The posterior distributions of this retrieval are shown in Figures~\ref{fig:K2-18b_multi} and \ref{fig:K2-18b_corner}. We obtain a precise and accurate estimate of the CH$_3$Cl abundance at $\mathrm{log}(X_{\rm CH_3Cl})$ = $-5.00^{+0.46}_{-0.55}$. This brings the CH$_3$Cl estimate in line with those obtained for the other biomarkers, except CS$_2$, as shown in Figure~\ref{fig:K2-18b_multi}. The retrieved spectrum and the simulated data for this case are shown in Figure~\ref{fig:spectra}.

We find that all five biomarkers in this scenario are  detectable in K2-18~b with a reasonable amount of JWST time. With our baseline configuration of one transit of K2-18~b with NIRISS and three with NIRSpec G395M, we find that DMS and OCS are detected at $\sim$4$\sigma$, while the remaining three biomarkers are detected at $\sim$2-3$\sigma$. Even with a total of only two transits, one each with NIRISS and NIRSpec G395M, we still detect DMS at 4$\sigma$ confidence. This is possible owing to NIRSpec G395M achieving the highest precision in the region where DMS has its strongest absorption peak, as can be seen in Figure~\ref{fig:spectra}. We have also considered a case including one transit with NIRISS and five with NIRSpec G395M, which is similar to observations that have been approved with these instruments in JWST Cycle 1 programs. For this configuration, we find that DMS is detected at over 6$\sigma$ and the remaining four biomarkers are all  detected at over 3$\sigma$. We therefore find that biomarkers are readily detectable in K2-18~b with JWST, although their detectability relies strongly on the abundances of the biomarkers and dominant species present, as well as the quality of observations. We predict that the approved Cycle 1 JWST observations of K2-18~b will be able to detect these biomarkers if present at the quantities considered here.

As seen in Figure~\ref{fig:K2-18b_wavcov}, we find that both NIRISS and NIRSpec G395M are necessary to obtain tight constraints on the abundances of both the dominant molecules and the trace biomarkers. We find that using only NIRISS data, thereby limiting the wavelength range to $\sim$1-2.8 $\mu$m, only yields constraints on the abundances of the dominant molecules. Conversely, using only NIRSpec G395M observations does not meaningfully constrain the abundances of the dominant molecules, while offering less precise constraints on the abundances of the five biomarkers compared to using both instruments together. Similarly, an underabundance of the biomarkers, below 1~ppmv, or overabundance of the prominent molecules CH$_4$ and NH$_3$ can affect the detectability of some biomarkers, particularly the ones with weaker or limited spectral features such as CH$_3$Cl and OCS. On the other hand, DMS is the most promising of all the biomarkers owing to its multiple strong features across the 1-5 $\mu$m range, making it readily detectable even with only two JWST transits, i.e., one each with NIRISS and NIRSpec G395M, as noted above. Furthermore, we find that the 2.9-5.1 $\mu$m range probed by NIRSpec G395M is necessary (but not sufficient) to constrain the abundances of all five biomarkers, due to both their multiple absorption bands in this range and the relative lack of strong features of the prominent molecules. To obtain the tightest constraints on biomarker abundances, we find it necessary to combine NIRSpec G395M observations with NIRISS which will also constrain the abundances of the prominent molecules.

\subsubsection{Case Studies: TOI-270~d and TOI-732~c} 
\label{sec::case_study2}

We investigate the potential for biomarker detection in two other Hycean planet candidates: TOI-270~d \citep{gunther2019} and TOI-732~c \citep{cloutier2020,nowak2020}. TOI-270~d has a radius of 2.01 $R_\oplus$ and a mass of 4.78 $M_\oplus$ \citep{VanEylen2021}. It orbits its host star, an M3V-type star, at a distance of 0.0722 au, giving it an equilibrium temperature of $T_{\rm eq}$ = 327~K. TOI-732~c has a radius of 2.42 $R_\oplus$ and a mass of 6.29 $M_\oplus$ \citep{nowak2020}. It orbits its M3.5V-type host at a semi-major axis of 0.0762 au and an eccentricity of 0.12, giving it a $T_{\rm eq}$ of 288~K and 324~K at its apocenter and pericenter, respectively. Table \ref{tab:planet_list} lists the full properties of the two planetary systems. With stellar masses in the range 0.38-0.39~$M_\odot$ and $T_{\rm eq}$ values well below $\sim$400~K, both planets are well within the Hycean HZ (see table \ref{tab:IHB}). The equilibrium temperatures can be higher for Bond albedos below the 0.5 value assumed here. 

For each of these two planets we generate a synthetic transmission spectrum using a nominal isothermal terminator temperature structure, set to 350~K for both TOI-270~d and TOI-732~c for illustration purposes. Following our results for K2-18~b above, we simulate the model atmospheres with the same abundances that yielded good constraints for all five biomarkers, i.e. $X_{\rm H_2O}$ = 10$^{-2}$, $X_{\rm CH_4}$ = 5$\times$10$^{-5}$, $X_{\rm NH_3}$ = 10$^{-5}$, 10 ppmv for CH$_3$Cl, and 1 ppmv for the other four biomarkers. We use the same instrument configurations as for K2-18~b described at the start of Section~\ref{sec::detectability}, allocating one transit for NIRISS observations and three transits for NIRSpec G395M. The resulting synthetic data for both planets, as well as the corresponding retrieved spectral fits, are shown in Figure~\ref{fig:spectra}. 

As shown in Figure~\ref{fig:planet_comp}, all five biomarkers are accurately constrained for both TOI-270~d and TOI-732~c, with CS$_2$ now also being precisely retrieved. For TOI-270~d, the retrieval yields biomarker estimates that are more precise than those for K2-18~b, with uncertainties of $\sim$0.3~dex for DMS, CS$_2$ and CH$_3$Cl and $\sim$0.5~dex for OCS. N$_2$O is retrieved with 0.6~dex and 1~dex upper and lower 1$\sigma$ uncertainties, respectively. The three dominant molecules are retrieved to within 0.3~dex, the only exception being the lower 1$\sigma$ uncertainty for NH$_3$ at 0.7~dex. In the case of TOI-732~c, the biomarker abundance values are again retrieved precisely, with even smaller uncertainties of $\sim$0.25~dex for DMS, CS$_2$ and CH$_3$Cl and $\sim$0.4~dex for OCS and N$_2$O. The three dominant molecules are all constrained to $\sim$0.3~dex or better. This is a consequence of their host stars being brighter than K2-18, leading to a higher spectroscopic precision. Additionally, their higher atmospheric temperatures yield larger scale heights and hence a larger signal-to-noise ratio compared to K2-18~b as seen in the synthetic spectra and corresponding retrieved spectra shown in Figure~\ref{fig:spectra}.

For both planets, all five biomarkers are retrieved at better precision and detected at greater significance compared to those for K2-18~b discussed above. For the same baseline instrument configuration for both planets (i.e. one transit with NIRISS and three transits with NIRSpec G395M), all 5 biomarkers are detected with a significance $\gtrsim$4-5$\sigma$, with the exception of N$_2$O in TOI-270~d which is detected at $>$2$\sigma$. We further find that despite the favorable conditions for such planets, reverting to canonical abundances yields a nondetection for CH$_3$Cl, as is the case with K2-18~b. However, the remaining four biomarkers are retrieved with precision comparable to or better than that of K2-18~b. 

Overall, our results show that the detection of all five biomarkers is possible under these conditions for a range of Hycean planets. Given that such detections are achievable for the Hycean planets shown here, we expect that biomarker detection is also possible for Dark Hycean planets, whose somewhat higher temperatures and, hence, larger scale heights can facilitate even more precise abundance estimates. However, if biomarker abundances are below 1 ppmv or there is a higher abundance of CH$_4$ and NH$_3$, we expect the detectability of biomarkers to vary on a case-by-case basis.

\section{Summary and Discussion}
\label{sec:discussion}

We investigate Hycean planets, a class of habitable planets with massive oceans and H$_2$-rich atmospheres. The internal structures of such planets lie between super-Earths that are dominated by rocky interiors and mini-Neptunes with H$_2$-rich envelopes too large to be habitable. We study the bulk properties (masses, radii, and temperatures), potential for habitability, and observable biosignatures of such planets. The wide range of conditions permissible on such planets make them conducive for detection, as well as atmospheric characterization, including the detection of biosignatures. Our study is motivated by the recent inference of the potential habitability of the exoplanet K2-18~b \citep{madhu2020}, which we now classify as a candidate Hycean world. 

Hycean planets span a significantly wider space in the mass-radius plane relative to habitable planets considered in previous studies. Across the range of habitable conditions considered in this work we find that Hycean planets can be as large as 2.6 $R_\oplus$ (2.3 $R_\oplus$) for a planet mass of 10 $M_\oplus$ (5 $M_\oplus$), with maximum equilibrium temperatures of $\sim$500 K. These limits assume that the planet has a rocky core in the interior that is at least 10\% by mass and is of Earth-like composition. These radii are significantly larger than those considered in the past for habitable Earth-like planets, as well as habitable ocean worlds \citep[e.g.,][]{leger2004,sotin2007,alibert2014} and habitable rocky super-Earths with H$_2$-rich atmospheres \citep[e.g.,][]{pierrehumbert2011,seager2013b}. As such, Hycean planets open a significantly wider discovery space in the search for potentially habitable planets. We identify a sample of promising Hycean candidates that are conducive for atmospheric characterization. Hycean planets also allow for a substantially wider HZ compared to the terrestrial HZ motivated by Earth-like conditions. 

We investigate the extent of the Hycean HZ for host stars ranging from late M dwarfs to sun-like stars. We find that the inner boundary of the regular Hycean HZ corresponds to $T_{\rm eq}$ as high as $\sim$430~K, depending on stellar type; higher $T_{\rm eq}$ correspond to cooler stars. For the outer boundary, Hycean planets can remain habitable for arbitrarily large orbital separations. In particular, Hycean planets can be habitable even with negligible or zero irradiation, as would be the case for planets on very large orbital separations and free-floating Hycean planets - we call these Cold Hycean worlds. Our finding for the outer HZ is consistent with that suggested for poorly irradiated or isolated rocky planets with thin H$_2$-rich atmospheres \citep{stevenson1999,pierrehumbert2011}. We also propose a further subclass of Hycean planets called Dark Hycean worlds, which are tidally locked planets with inefficient day-night energy redistribution whose permanent nightsides could be habitable even if the dayside is too hot. Such planets could have  a planet-wide average $T_{\rm eq}$ up to 510~K, or higher, and still be habitable on the nightside depending on the albedo and day-night energy redistribution.  

We investigate the detectability of biomarkers in the atmospheres of Hycean worlds. The dominant gases in Hycean atmospheres, besides H$_2$/He, may be expected to be H$_2$O, CH$_4$ and NH$_3$, all of which are expected to be naturally occurring in chemical equilibrium abiogenically. We note, however, that CH$_4$ and NH$_3$ can be depleted due to disequilibrium processes \citep[e.g.,][]{madhu2020, Yu2021} The primary biomarkers in terrestrial-like atmospheres such as O$_2$/O$_3$ and CH$_4$ \citep[e.g.,][]{catling2018} are expected to be underabundant and/or abiogenic in H$_2$-rich atmospheres.  However, we consider several secondary terrestrial biomarkers that may be expected to be present in trace quantities ($\sim$1 ppmv) in oceanic environments with life, e.g., DMS, CS$_2$, CH$_3$Cl, OCS, and N$_2$O \citep{segura2005,domagal-goldman2011, seager2013a,seager2016}. 

We find that all these biomarkers are detectable in nearby transiting Hycean atmospheres using transmission spectroscopy with modest amount of JWST time. We conduct atmospheric retrievals on simulated spectra of three candidate Hycean planets and demonstrate accurate abundance estimates of the biomarkers to precisions smaller than $\sim$1~dex and as low as $\sim$0.25~dex for realistic atmospheric compositions. Our results agree with previous studies which suggested that such biomarkers can be detected in atmospheres of rocky exoplanets with H$_2$-rich atmospheres observed with JWST \citep{seager2013b}. We find that the larger radii and higher temperatures admissible for Hycean planets make these biomarkers more readily detectable in Hycean atmospheres compared to those of rocky exoplanets. In particular, we predict that the approved Cycle 1 JWST observations of K2-18~b, a candidate Hycean planet, will be able to detect these biomarkers if present at the quantities considered in this work.

\subsection{Factors Affecting Habitability}
Following many previous studies, we have defined the HZ based on the requirement of liquid water at the planetary surface \citep[e.g.,][]{kasting1993,kasting2003,selsis2008,forget2013,kaltenegger2017,kopparapu2018,meadows2018book}, with the additional requirement of surface temperatures known to be habitable on Earth \citep{rothschild2001,merino2019}. However, other physical factors are also involved in determining habitability. One such factor is the role of geochemical cycling in regulating atmospheric composition and surface temperature, e.g. the carbonate-silicate cycle on Earth \citep[e.g.,][]{walker1981,kasting1993,franck2000,lammer2010}. While this has been widely studied in the context of Earth and terrestrial planets, such cycles would evidently be very different for Hycean planets. Future work will be needed to establish how such processes work.

Another significant factor affecting habitability is stellar activity and stellar winds \citep[e.g.,][]{khodachenko2007,lammer2007,rodriguez-mozos2019}. This is especially relevant for M-dwarf planets, as these stars are known to be more active than hotter stars \citep[e.g.,][]{shields2016}. UV flux, coronal mass ejections, and stellar winds can gradually erode planetary atmospheres and potentially damage life existing on the surface \citep[but see, e.g.][]{omalley-james2017,omalley-james2019}. However, more massive planets may be more robust to stellar activity owing to factors such as higher gravity, stronger magnetic moments, and thicker atmospheres \citep[e.g.,][]{lammer2007,kopparapu2014}. Planets with thicker atmospheres could also plausibly limit the UV flux reaching their surfaces, thereby protecting any existing life. In the context of stellar activity, Hycean planets orbiting M-dwarf hosts may therefore provide better chances for habitability compared to terrestrial planets in similar conditions. 

Habitability also requires the maintenance of liquid surface water for a significant period of time such that life can be initiated and subsequently sustained. In the case of terrestrial-like planets, water loss at the inner edge of the HZ can preclude life by quickly removing the planetary water reservoir, especially around active stars \citep[e.g.,][]{luger2015b,wolf2015,ribas2016,bolmont2017,kopparapu2017,wordsworth2018}. However, for Hycean planets, the planetary water reservoir is very large (over 10\% by mass), and water is unlikely to be exhausted by atmospheric escape. This also allows for higher temperatures at the ocean surface, up to ~400 K or higher, without the risk of total runaway loss of the ocean. A further consideration for the maintenance of liquid water is orbital dynamics. For example, a highly eccentric or otherwise perturbed orbit may change the irradiation incident on the planet on fairly short timescales and may therefore preclude the stability of liquid surface water \citep[e.g.,][]{dvorak2010,kopparapu2010,bolmont2016,palubski2020}.

\subsection{Future Prospects} 

Some of the challenges underlying the characterization of habitable rocky exoplanets are also common to Hycean planets. First, while mass and radius are imperative to establish whether a certain planet is a Hycean candidate (see, e.g., Figure~\ref{fig:mr}), they are not sufficient to confirm a unique interior composition due to natural compositional degeneracies \citep[e.g.,][]{rogers2010,madhu2020}. Second, even if a candidate Hycean planet is in the Hycean HZ it may not necessarily have the right conditions for habitability, e.g., the internal structure and atmospheric properties may be such that the ocean surface pressure and/or temperature is too high. Finally, the detection of H$_2$O in the atmosphere does not guarantee the presence of an ocean on the planet, as H$_2$O can be naturally occurring in H$_2$-rich atmospheres as the prominent oxygen-bearing species. Conversely, the nondetection of H$_2$O does not preclude the presence of an ocean, since at low atmospheric temperatures H$_2$O can rain out and not be detectable in the atmosphere. Nevertheless, in all these aspects Hycean candidates offer better prospects for establishing their habitability compared to habitable rocky exoplanets, which are inherently harder to characterize. 

Observationally, Hycean planets provide a promising avenue in the search for habitable exoplanets and their biosignatures. Demographics of exoplanetary systems discovered by transit surveys \citep[e.g.,][]{fulton2018,hardegree-ullman2020} show that the known exoplanet radius distribution peaks in the Hycean range between 1 and 2.6 $R_\oplus$. Thus, Hycean worlds could potentially be ubiquitous in nature. Hycean planets are also optimal targets for atmospheric spectroscopy of habitable planets using current and future facilities. Habitable rocky exoplanets with heavy molecular atmospheres (e.g., of H$_2$O, CO$_2$, N$_2$, or O$_2$) are expected to have small scale heights, making them challenging for atmospheric spectroscopy. For example, detection of biomarkers on rocky exoplanets such as TRAPPIST-1~d could require tens of transits with JWST \citep{barstow2016,lustig-yaeger2019}. On the other hand, H$_2$-rich atmospheres with larger scale heights are more favorable for atmospheric characterization. The potential for biosignature detection in H$_2$-rich atmospheres of rocky exoplanets has already been suggested \citep[e.g.,][]{seager2013b}. The prospects of such biomarker detections are even more favorable for a Hycean planet, which has not only an H$_2$-rich atmosphere but also a substantial H$_2$O ocean underneath, potentially providing a large biosphere. The combination of large radii and large atmospheric scale heights makes Hycean planets optimal targets for atmospheric spectroscopy. 

We hope our study provides impetus in expanding the search for habitable planets and biosignatures beyond the conventional boundaries of rocky exoplanets. Such an effort could bring the search for biosignatures within the reach of upcoming facilities in the near future.   

\section*{Acknowledgements}
We thank the anonymous reviewer for their valuable comments and the Editorial team at AAS journals for efficiently overseeing the review and publication of our work during the challenging past year of the COVID-19 pandemic. We thank all those in Cambridge and beyond who worked on the frontlines to keep us safe during the pandemic. A.A.A.P. acknowledges support from the UK Science and Technology Facilities Council (STFC) toward her doctoral studies. We thank Siddharth Gandhi for discussion on day–night energy redistribution in the self-consistent atmospheric models, Matthew Nixon for discussion on the water equation of state, and Subhajit Sarkar for discussion on JWST simulated data. We thank James Kasting for helpful feedback on our manuscript. This research has made use of the NASA Exoplanet Archive, which is operated by the California Institute of Technology, under contract with the National Aeronautics and Space Administration under the Exoplanet Exploration Program. This research has made use of the NASA Astrophysics Data System and the Python packages \textsc{numpy}, \textsc{scipy} and \textsc{matplotlib}. Part of this work was performed using resources provided by the Cambridge Service for Data Driven Discovery (CSD3) operated by the University of Cambridge Research Computing Service (\url{www.csd3.cam.ac.uk}), provided by Dell EMC and Intel using Tier-2 funding from the Engineering and Physical Sciences Research Council (capital grant EP/P020259/1), and DiRAC funding from the Science and Technology Facilities Council (\url{www.dirac.ac.uk}).

\newpage

\appendix

\section{Habitable Zone Calculations}
\label{sec:star_props}

Table \ref{tab:star_props} shows the stellar properties used in section \ref{sec:hycean_hz}, as well as the prototype stars they are based on.

\begin{table}[htbp]
  \caption{Stellar Properties Assumed in This Work and the Prototype Stars They Are Based On.}
  
  \centering
    \begin{tabular}{*{6}{c|}c}
    \hline
     $T_\star$/K & $ M_\star/M_\odot$ & $R_\star/R_\odot$ & log($ g$/cm~s$^{-2}$) & [Fe/H] & Prototype &  Ref \\
    \hline
    \hline
    2500  & 0.08  & 0.12 & 5.0     & 0.0  & TRAPPIST-1 & 1 \\
    3000  & 0.12 & 0.14 & 5.0 & 0.0  & Proxima Cen & 2 \\
    3000  & 0.16 & 0.21 & 5.0 & 0.5  & GJ 1214 & 3,4 \\
    3300  & 0.26 & 0.28 & 5.0 & -0.5 & LTT 1445 A & 5 \\
    3400  & 0.31 & 0.31 & 4.9 & 0.0 & TOI-175 & 6 \\
    3590  & 0.44 & 0.45 & 4.9 & 0.1 & K2-18 & 7 \\
    4145  & 0.58  & 0.57 & 4.6   & -0.1 & WASP-80 & 8 \\
    4430  & 0.69  & 0.66  & 4.5   & 0.0  & WASP-107& 9 \\
    4750  & 0.80   & 0.74  & 4.6  & 0.2  & WASP-132 & 10 \\
    5275  & 0.93  & 0.87  & 4.5   & 0.0    & CoRoT-7 & 11 \\
    5777  & 1.00  & 1.00 & 4.4 & 0.0    & Sun & 12 \\
    6025  & 1.18  & 1.38  & 4.2  & 0.1  & K2-236 & 13 \\
    \hline
    \end{tabular}
    \begin{tablenotes}
    \item {\bf Note:} For each star, we use either a Phoenix model (for 2500 $\leq T_\star \leq $ 3500~K) or a Kurucz model (for $T_\star >$ 3500~K) for the stellar spectrum assuming the gravity (log($g$)), [Fe/H] metallicity, and effective temperature ($T_\star$) listed (see section \ref{sec:atmos_model}). The values shown here are based on values used in the references listed for each planet-hosting prototype star.\newline
    \footnotesize{ {\bf References:}
    (1)~\citet{gillon2017}; (2)~\citet{anglada-escude2016}; (3)~\citet{charbonneau2009}; (4)~\citet{rojas2010}; (5)~\citet{winters2019}; (6)~\citet{cloutier2019b};  
    (7)~\citet{hardegree-ullman2020}; (8)~\citet{triaud2013}; (9)~\citet{anderson2017}; (10)~\citet{hellier2017}; (11)~\citet{leger2009}; (12)~\citet{cox2000}; (13)\citet{chakraborty2018}.
    }    
    \end{tablenotes}
  \label{tab:star_props}
\end{table}

\begin{figure}
    \centering
    \includegraphics[width=0.37\textwidth]{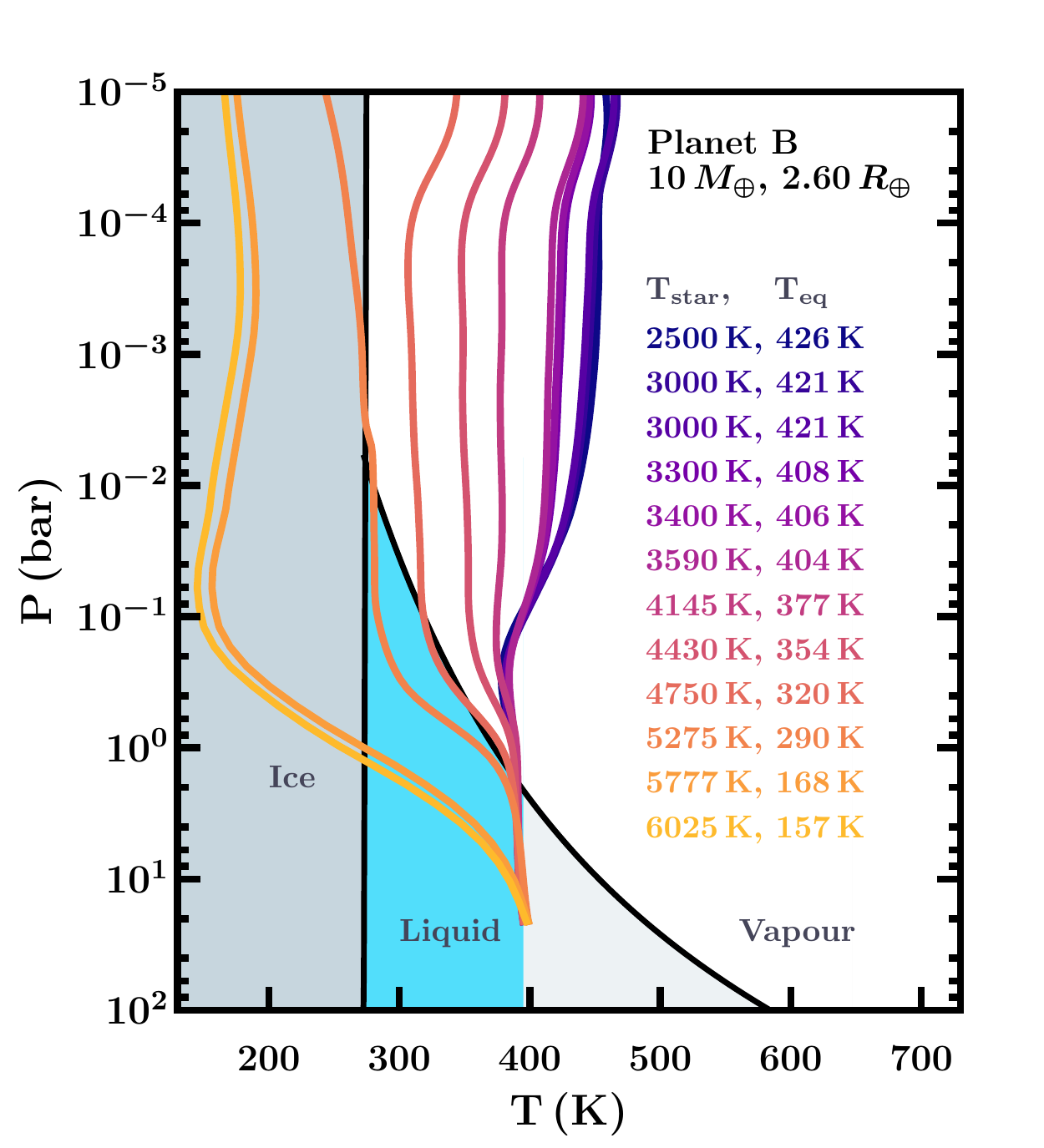} \hspace{-0.1cm}
    \includegraphics[width=0.62\textwidth]{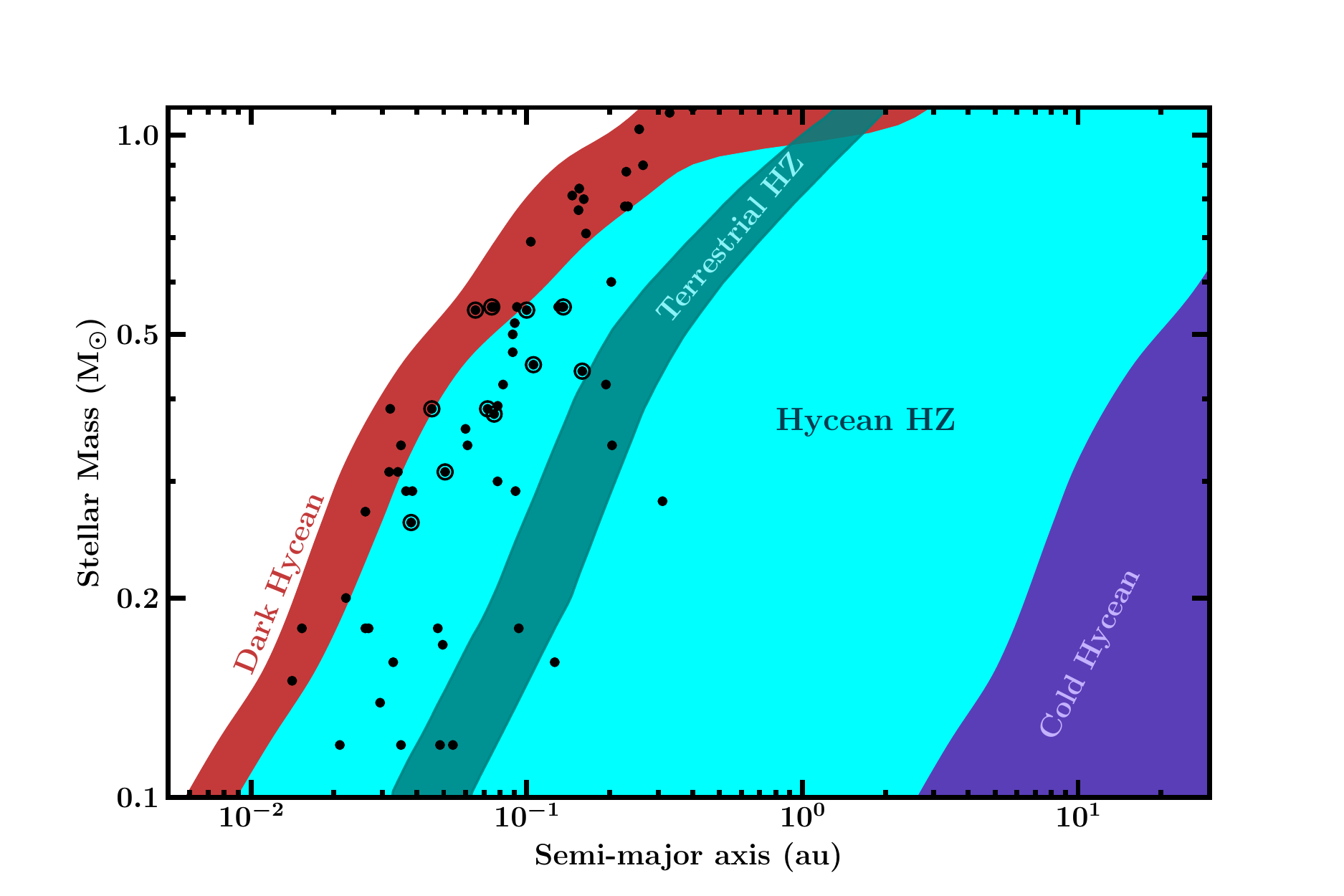}
    \caption{\emph{Left:} Temperature profiles for Planet B as in the right panel of Figure \ref{fig:dayside_fixedAb} but with $P_{\rm HHB}$~=~21~bar such that the atmospheric H$_2$O is 100\% saturated near the HHB. \emph{Right:} The Hycean HZ as in Figure~\ref{fig:habzone} but for the $P_{\rm HHB}$~=~21~bar case with 100\% saturation.}
    \label{fig:appendixA_hz}
\end{figure}

In section~\ref{sec:hycean_hz}, we discuss the temperature structures and inner HZ for the limiting case with P$_{\rm HHB}$~=~2.1 bar, T$_{\rm HHB}$~=~395 K and 10\% saturation of atmospheric H$_2$O near the HHB. Here, we consider a case with 100\% saturation at the HHB, obtained for P$_{\rm HHB}$ = 21 bar with the same T$_{\rm HHB}$ of 395 K and the same atmospheric abundances as in Section \ref{sec:hab}. We consider planet B for this case as pursued in section~\ref{sec:hycean_hz}. The temperature profiles for this set up are shown in the left panel of Figure \ref{fig:appendixA_hz}, and have similar values of $T_{\rm eq}$ corresponding to the IHB compared to the P$_{\rm HHB}$~=~2.1 bar case (right panel of Figure \ref{fig:dayside_fixedAb}). We also find that for this case the Dark Hycean IHB occurs at $T_{\rm eq,av}$~=~525~K, which is close to the 511~K limit we find in Section \ref{sec:dark_hycean_hz}. The Hycean and Dark Hycean HZs for this case are shown in the right panel of Figure \ref{fig:appendixA_hz}, and are similar to those in Figure \ref{fig:habzone} for the P$_{\rm HHB}$~=~2.1 bar case.

\section{Treatment of Day-Night Flux Redistribution}
\label{sec:appendix_redist}

In our nightside atmospheric models in section \ref{sec:dark_hycean_hz}, we account for energy flux advected from the dayside to the nightside. This is performed by adding an energy source in the equation of radiative equilibrium, as described in \citet{burrows2008}. The \textsc{Genesis} atmospheric model solves both the integral and differential forms of the radiative equilibrium equation, in different parts of the atmosphere. The differential form,
\begin{equation}
    \label{eq:diff_form}
    \int_{0}^{\infty}\frac{\mathrm{d}(f_{\nu}J_{\nu})}{\mathrm{d}\tau_\nu}\mathrm{d}\nu = \frac{\sigma_\mathrm{\textsc{sb}}}{4\pi}T_\mathrm{int}^4,
\end{equation}
is relevant in the deeper regions, where the optical depth, $\tau_\nu$, is large. Here, $J_{\nu}=\frac{1}{2}\int_{-1}^{1}I_\nu(\mu)d\mu$, $f_{\nu}=\frac{1}{2}\int_{-1}^{1}\mu^2I_\nu(\mu)d\mu/J_\nu$ and $I_\nu(\mu)$ is the specific intensity. $\frac{\sigma_\mathrm{\textsc{sb}}}{4\pi}T_\mathrm{int}^4$ represents the net internal flux emanating from the interior of the planet, where $T_\mathrm{int}$ is the internal temperature and $\sigma_\mathrm{\textsc{sb}}$ is the Stefan-Boltzmann constant. This form is required at deeper pressures to set the net level of outgoing flux, but is numerically unstable at lower pressures when $\mathrm{d}\tau_\nu$ becomes small. Therefore, at lower pressures, the integral form is used:
\begin{equation}
    \label{eq:int_form}
    \int_{0}^{\infty} \kappa_\nu (J_\nu-B_\nu)\mathrm{d}\nu = 0.
\end{equation}
$\kappa_\nu$ is the absorption coefficient, and $B_\nu$ is the Planck function evaluated at the temperature of a given atmospheric layer. Note that in equations \ref{eq:diff_form} and \ref{eq:int_form}, we do not include terms due to convection for clarity. In convective regions, these equations are modified to include convective flux as described in \citet{gandhi2017}.

We assume that the day-night redistributed flux is advected across a given pressure range. From the bottom to the top of this pressure range, the redistributed flux incrementally adds to the net outgoing flux such that at the top of the atmosphere the total net flux is $\frac{\sigma_\mathrm{\textsc{sb}}}{4\pi}T_\mathrm{int}^4+H_\mathrm{irr}$. $H_\mathrm{irr}$ is the total flux transported from the dayside to the nightside, expressed as the $H$-moment (i.e. flux/$4\pi$). Assuming a dayside irradiation temperature $T_\mathrm{irr}$ and a redistribution efficiency $P_\mathrm{n}$ (using the notation of \citet{burrows2008}), 
\begin{equation*}
    H_\mathrm{irr} = P_\mathrm{n} \frac{\sigma_\mathrm{\textsc{sb}}}{4\pi}T_\mathrm{irr}^4.
\end{equation*}
Equations \ref{eq:diff_form} and \ref{eq:int_form} are modified to account for this flux as follows:
\begin{equation*}
    \label{eq:diff_form_redist}
    \int_{0}^{\infty}\frac{\mathrm{d}(f_{\nu}J_{\nu})}{\mathrm{d}\tau_\nu}\mathrm{d}\nu = \frac{\sigma_\mathrm{\textsc{sb}}}{4\pi}T_\mathrm{int}^4 + H_\mathrm{irr},
\end{equation*}
\begin{equation*}
    \label{eq:int_form_redist}
    \int_{0}^{\infty} \kappa_\nu (J_\nu-B_\nu)\mathrm{d}\nu = -D(z),
\end{equation*}
where
\begin{equation*}
    \int_{z_\mathrm{min}}^{z_\mathrm{max}}D(z) \mathrm{d}z =  H_\mathrm{irr}
\end{equation*}
and $z_\mathrm{min}$, $z_\mathrm{max}$ are the minimum and maximum altitudes in the atmospheric model, respectively. $D(z)$ therefore sets the vertical profile of the redistributed flux. 

\begin{figure*}
    \centering
    \includegraphics[width=0.8\textwidth]{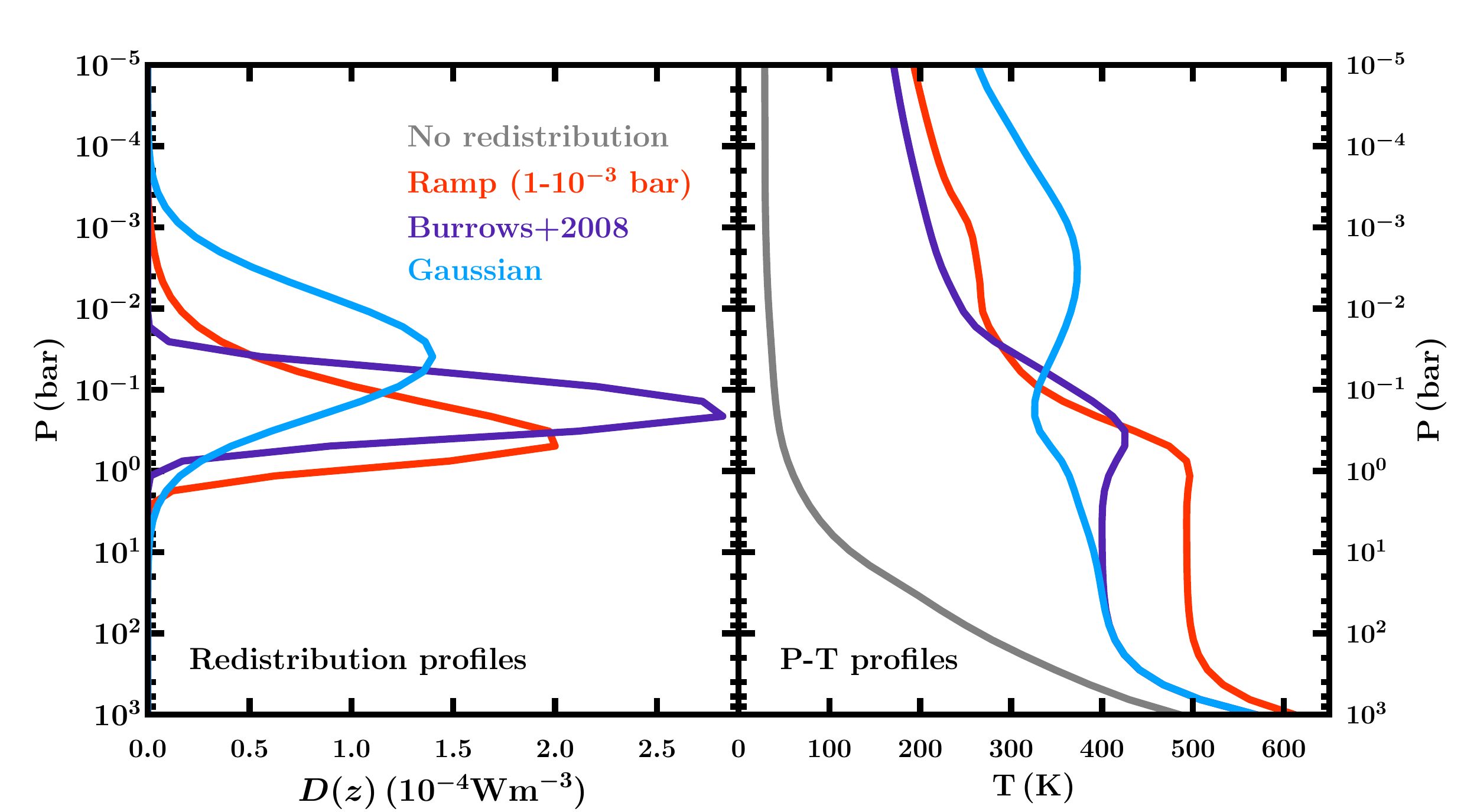}
    \caption{Left: redistribution profiles for the ramp and Gaussian cases described in equations \ref{eq:ramp} and \ref{eq:gaussian}, respectively. The red and purple lines show the ramp profile applied in the ranges 1-$10^{-3}$~bar and 0.5-0.05~bar (as in \citealt{burrows2008}), respectively. The blue line shows the Gaussian profile, whose upper and lower 2$\sigma$ intervals coincide with pressures of 1~bar and 1~mbar, respectively. All profiles assume $P_\mathrm{n}$~=~0.5 and a dayside $T_\mathrm{irr}$ of 400~K. Right: nightside $P$-$T$ profiles corresponding to each redistribution profile, assuming $T_\mathrm{int}$~=~25~K. The $P$-$T$ profile corresponding to no redistribution is shown in gray.}
    \label{fig:redist_methods}
\end{figure*}

Here, we consider two different functional forms for $D(z)$. Firstly, we consider the form used by \citet{burrows2008} (`model 2' in their appendix A), in which $D$ decreases linearly with surface density, $m$, between two limiting altitudes:
\begin{equation}
    \label{eq:ramp}
    D(m) =
    \begin{cases}
     \dfrac{2 H_\mathrm{irr}}{m_1-m_0}\dfrac{m_1-m}{m_1-m_0}, & \text{if } m_0<m<m_1\\
    0,& \text{otherwise}
    \end{cases}
\end{equation}
where $m_1$ and $m_0$ are the surface density at the lower and higher limiting altitudes, respectively. Then, since we require that $D(m)\mathrm{d}m=D(z)\mathrm{d}z$, $D(z)=\rho D(m)$, where $\rho$ is density and d$m$=$\rho$d$z$. We refer to this as the `ramp' model.

Secondly, we consider a Gaussian profile in log pressure:
\begin{equation}
    \label{eq:gaussian}
    D(\log(P)) = \dfrac{H_\mathrm{irr}}{\sigma\sqrt{2\pi}}\text{exp}\left(-\frac{1}{2}\left(\frac{\log(P)-\mu}{\sigma}\right)^2\right),
\end{equation}
and $D(z)=D(\log(P)) \rho g/P$. $\mu$ and $\sigma$ are the mean and standard deviation of the distribution in log pressure, respectively.

\begin{figure*}
    \centering
    \includegraphics[width=0.8\textwidth]{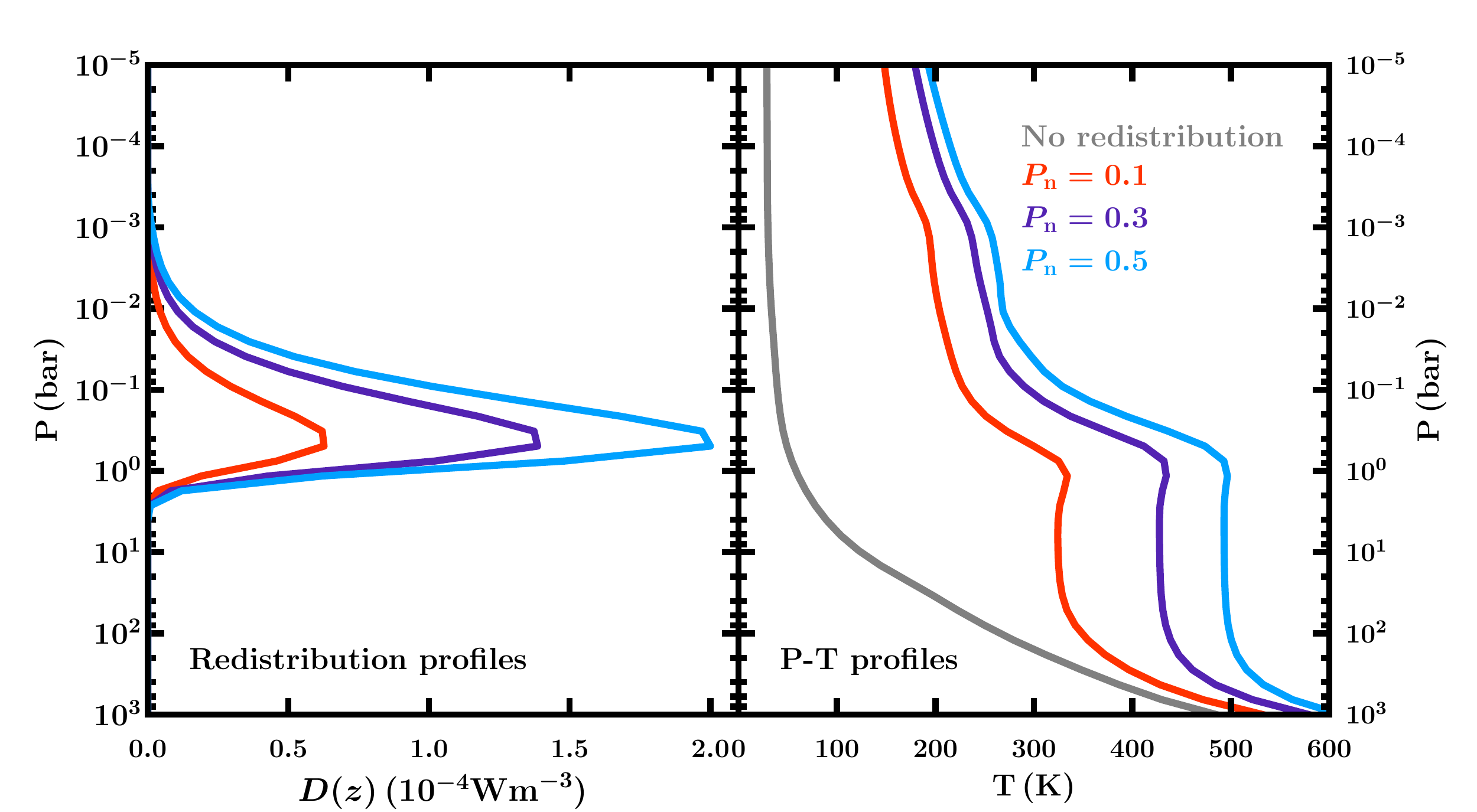}
    \caption{Left: ramp redistribution profiles (equation \ref{eq:ramp}) assuming $P_\mathrm{n}$=0.1 (red), 0.3 (purple), and 0.5 (blue). All profiles assume redistribution in the range 1-$10^{-3}$~bar and a dayside $T_\mathrm{irr}$ of 400~K. Right: nightside $P$-$T$ profiles corresponding to each redistribution profile, assuming $T_\mathrm{int}$=25~K. $P$-$T$ profile corresponding to no redistribution is shown in gray.}
    \label{fig:redist_Pn}
\end{figure*}

Figure \ref{fig:redist_methods} shows examples of the ramp and Gaussian redistribution profiles from equations \ref{eq:ramp} and \ref{eq:gaussian}, respectively. For the ramp profile, we show cases with different pressure ranges: 0.5-0.05~bar, as in \citet{burrows2008}, and 1-$10^{-3}$~bar. For the Gaussian profile, we place the mean of the distribution at 3$\times 10^{-2}$~bar, and use a standard deviation of 0.75 dex in pressure such that the 2$\sigma$ intervals occur at 1 and $10^{-3}$~bar. For all three profiles, we use $P_\mathrm{n}$~=~0.5 and $T_\mathrm{irr}$~=~400~K. Figure \ref{fig:redist_methods} also shows the corresponding nightside $P$-$T$ profiles for each redistribution profile, which are discussed below. Figure \ref{fig:redist_Pn} shows the effect of changing $P_\mathrm{n}$ on both the redistribution and $P$-$T$ profiles for the ramp model applied in the range 1-$10^{-3}$~bar.

In order to compare these redistribution profiles, we consider the pressures at which they transport flux. For both the ramp profile applied in the pressure range 1-$10^{-3}$~bar and the Gaussian profile, flux is largely redistributed within the same pressure range. However, within this range, the Gaussian profile redistributes a larger proportion of flux at lower pressures. Based on GCMs, it is known that energy redistribution tends to be more efficient at relatively deeper pressures, resulting in more homogeneous day-night temperature distributions at higher pressures \citep[e.g.,][]{showman2009}. This is consistent with the fact that density and temperature are typically higher at deeper pressures, increasing the efficiency of advection of energy from the day to the nightside. As a result, the ramp profile is a more physical representation of flux transport in the atmosphere, and we choose to use it in this work. 

The effects of each redistribution profile on the nightside atmospheric $P$-$T$ profile are shown in Figures \ref{fig:redist_methods} and \ref{fig:redist_Pn} for $T_\mathrm{int}$~=~25~K. As expected, the ramp redistribution profile results in more heat deposition at higher pressures relative to a Gaussian profile with the same pressure range. The \citet{burrows2008} model redistributes flux in an intermediate pressure range, which is reflected in the $P$-$T$ profile. Also as expected, Figure \ref{fig:redist_Pn} shows that as $P_\mathrm{n}$ is increased, the nightside $P$-$T$ profile becomes hotter. Furthermore, the presence of any redistribution significantly increases the temperature of the nightside at pressures $\lesssim$1000~bar compared to a model with no redistribution (shown in gray).

\bibliography{references}{}
\bibliographystyle{aasjournal}


\end{document}